\documentclass[twocolumn]{aastex63}
\newcommand{\code}[1]{\texttt{#1}}
\newcommand{\mesa}{\code{MESA}}

\newcommand\logten{\ensuremath{\log_{10}}}

\usepackage{booktabs}

\shorttitle{Type IIb SN Progenitors - II}
\shortauthors{Sravan et al.}

\begin{document}


\title{Progenitors of Type II\MakeLowercase{b} Supernovae: II. Observable Properties}


\author{Niharika Sravan}
\affiliation{Department of Physics and Astronomy, Purdue University, 525 Northwestern Ave., West Lafayette, IN 47907, USA}
\affiliation{Center for Interdisciplinary Exploration and Research in Astrophysics (CIERA)\\ and\\
Department of Physics and Astronomy, Northwestern University, 2145 Sheridan Road, Evanston, IL 60208, USA}
\author{Pablo Marchant}
\affiliation{Institute of Astrophysics, KU Leuven, Celestijnenlaan 200 D, 3001 Leuven, Belgium}
\author{Vassiliki Kalogera}
\affiliation{Center for Interdisciplinary Exploration and Research in Astrophysics (CIERA)\\ and\\
Department of Physics and Astronomy, Northwestern University, 2145 Sheridan Road, Evanston, IL 60208, USA}
\author{Dan Milisavljevic}
\affiliation{Department of Physics and Astronomy, Purdue University, 525 Northwestern Ave., West Lafayette, IN 47907, USA}
\author{Raffaella Margutti}
\affiliation{Center for Interdisciplinary Exploration and Research in Astrophysics (CIERA)\\ and\\
Department of Physics and Astronomy, Northwestern University, 2145 Sheridan Road, Evanston, IL 60208, USA}




\begin{abstract}
Type IIb supernovae (SNe IIb) present a unique opportunity for investigating the evolutionary channels and mechanisms governing the evolution of stripped-envelope SN progenitors due to a variety of observational constraints available.
Comparison of these constraints with the full distribution of theoretical properties not only help ascertain the prevalence of observed properties in nature, but can also reveal currently unobserved populations.
In this follow-up paper, we use the large grid of models presented in \citet{2019ApJ...885..130S} to derive distributions of single and binary SNe IIb progenitor properties and compare them to constraints from three independent observational probes: multi-band SN light-curves, direct progenitor detections, and X-ray/radio observations.
Consistent with previous work, we find that while current observations exclude single stars as SN IIb progenitors, SN IIb progenitors in binaries can account for them.
We also find that the distributions indicate the existence of an unobserved dominant population of binary SNe IIb at low metallicity that arise due to mass transfer initiated on the Hertzsprung Gap.
In particular, our models indicate the existence of a group of highly stripped (envelope mass $\sim 0.1- 0.2 M_\sun$) progenitors that are compact ($<50 R_\sun$) 
and blue ($T_{\rm eff} \lesssim 10^5$K) with $\sim10^{4.5}-10^{5.5}~L_\sun$ and low density circumstellar mediums.
As discussed in \citet{2019ApJ...885..130S}, this group is necessary to account for SN IIb fractions and likely exist regardless of metallicity.
The detection of the unobserved populations indicated by our models would support weak stellar winds and inefficient mass transfer in SN IIb progenitors.
\end{abstract}





\section{Introduction} \label{s:intro}
SNe are classified into Types I or II depending on the absence or presence of hydrogen emission lines in their spectra, respectively.
Except SNe Ia, that are thermonuclear explosions resulting from runaway fusion in white-dwarfs, 
Type I SNe result from iron core-collapse (CC) of massive stars that have lost their outer hydrogen layers. 
Type IIb SNe are an interesting class that transition from Type II to I, initially exhibiting prominent hydrogen spectral features that weaken and disappear over time.
They are related to Type I CC SNe except that their progenitor stars have lost most but not all of their hydrogen layers. 
These SNe are therefore collectively also referred to as stripped-envelope (SE) SNe.

A key question about SE SNe is the mechanisms that drive the removal of the envelopes of their progenitor stars. 
In particular, the debate centers around the relative roles, if any, of stellar winds \citep[e.g.,][]{1993ApJ...411..823W,2012A&A...542A..29G,2013A&A...558A.131G}, stellar rotation \citep[e.g.,][]{2012A&A...542A..29G,2013A&A...550L...7G,2013A&A...558A.131G,2020MNRAS.tmp.1240Z}, binary interactions \citep[e.g.,][]{1992ApJ...391..246P,2010ApJ...725..940Y,2017ApJ...840...10Y, 2017MNRAS.470L.102S,2019MNRAS.490....9L}, and, nuclear burning instabilities \citep[e.g.,][]{2011ApJ...741...33A,2015ApJ...811..117S}. 
There has been growing support for binary interactions as dominant due to several independent lines of evidence, including weaker stellar winds \citep{2014ARA&A..52..487S} and higher binary fractions \citep{2012Sci...337..444S, 2017ApJS..230...15M} than previously estimated.
Other key lines of support include high SE SN fractions \citep{2011MNRAS.412.1522S}, low ejecta masses \citep{2016MNRAS.457..328L,2019MNRAS.485.1559P}, and, low circumstellar medium (CSM) densities \citep{2012ApJ...752...17W,2016ApJ...821...57D}.

SNe IIb in particular are important for addressing which stripping mechanisms dominate because they have five identified progenitors, compared to just one for each of the other SE SN types. 
There is also evidence for a binary companion to the progenitor star in three cases.
These are in addition to several constraints on progenitor structure from analyzing multi-band SN light-curves (LCs) and CSM properties from X-ray/radio observations (see Table \ref{t:SNIIbprops}). 

Several theoretical investigations have leveraged this unique opportunity to constrain the mechanisms governing the evolution of SNe IIb progenitors.
\citet{2011A&A...528A.131C} used a large grid of solar metallicity single and binary SN IIb models\footnote{\citet{2011A&A...528A.131C} defined SN IIb progenitors as those that explode with $0.1-0.5 M_\sun$ of residual hydrogen envelope.} 
to identify various evolutionary channels resulting in SNe IIb and their observational characteristics, focusing on the progenitor's and companion's photometric properties. However, they did not investigate CSM properties.
\citet{2017ApJ...840...10Y} used a broad but sparse grid of binary models with fixed mass ratio and mass transfer efficiency at solar and sub-solar metallicities to examine the structural, photometric, and CSM properties of SN IIb progenitors.
\citet{2017ApJ...840...90O} focused on investigating CSM properties of SN IIb progenitors using a small grid of solar metallicity binary models.

Though important for delineating the range of progenitor properties, constraints from limited or sparse model grids do not convey full information.
For example, \citet{2004Natur.427..129M} favored a binary SN IIb progenitor evolving via case C mass transfer (i.e. mass transfer after core helium exhaustion) to explain photometric and spectroscopic constraints for SN 1993J's progenitor and companion.  
However, this mass transfer scenario is only possible for a small set of binary configurations where the progenitor avoids mass transfer during its first larger rise on the giant branch, while initiating mass transfer on its second smaller rise after core helium exhaustion \citep{2017ApJ...840...10Y,2019ApJ...885..130S}.
On the other hand, a large grid of models can be used to derive probability distributions of progenitor properties by accounting for the probability of existence of a given progenitor in nature.
In addition to showing the range of properties, probability distributions also show the statistical significance of a given progenitor and its associated observational characteristics.
This is especially important as we increasingly build larger observational samples facilitating derivation of probability distributions of observed properties \citep{2016MNRAS.457..328L,2018A&A...609A.136T,2019MNRAS.485.1559P}.
Moreover, theoretical probability distributions of progenitor properties can also help guide observational searches, to either validate or falsify models, using observational regimes favored theoretically but absent in existing observations.

In this second paper of a two-part study aimed at conducting a comprehensive investigation of SN IIb progenitors, we use an extensive grid of single and binary SN IIb models to
investigate the distribution of observable properties of SNe IIb constrained using three independent observational probes: analyzing LCs, direct progenitor detections, and X-ray/radio observations. 
We identify regions of disagreement with current observational constraints and assess whether this is could be due to current observational bias or the corresponding evolutionary channels not being realized in nature.
This paper builds on the conclusions from the first part \citep[][henceforth referred to as paper I]{2019ApJ...885..130S}, where we examined the parameter space, evolutionary pathways and fractions for SNe IIb, 
to create a comprehensive picture of our theoretical understanding of their progenitors.

This paper is organized as follows. 
In Section \ref{s:models} we summarize our models and methods.
In Section \ref{s:relations} we examine correlations between physical properties of our models. 
In Section \ref{s:obs} we discuss the three available observational probes into SNe IIb progenitors (from analyzing LCs, direct detections, and X-ray/radio observations) and their limitations.
In Section \ref{s:results} we compare our models to constraints from these probes for all SNe IIb analyzed in the literature and highlight regions of (dis)agreement.
In Section \ref{s:oppor} we delineate currently unobserved observational regimes that could constrain SN IIb progenitor channels.
We summarize and conclude in Section \ref{s:conclusions}.

\section{SN II\lowercase{b} Models} \label{s:models}

In this section, we briefly summarize our models. These models are same as those in paper I. We refer the reader to Section 2 therein for detailed discussions. All models including \mesa\ input files used to produce them are available at \url{https://doi.org/10.5281/zenodo.3332830}. 

We model non-rotating single and binary star models at solar ($Z_\sun= 0.02$) and sub-solar (1/4 $Z_\sun$; henceforth referred to as `low') metallicities using Modules for Experiments in Stellar Astrophysics \citep[\mesa\ Release 9575;][]{2011ApJS..192....3P,2013ApJS..208....4P,2015ApJS..220...15P,2018ApJS..234...34P}. We assume that the helium abundance increases linearly from Y = 0.2477 \citep{2007ApJ...666..636P} at Z = 0.0 to Y = 0.28 at Z = 0.02 \citep{2011A&A...530A.115B}. 
We use the \texttt{basic.net}, \texttt{co\_burn.net}, and \texttt{approx21.net} nuclear networks in \mesa. 
Radiative opacities are computed using tables from the OPAL project \citep{1996ApJ...464..943I}. 

We model convection using the standard mixing-length theory (MLT), adopting the Ledoux criterion, with the mixing length parameter, $\alpha_{\rm MLT}$, set to 1.5.
We adopt the MLT++ prescription of \mesa\ which enhances energy transport in convective regions that approach the Eddington limit \citep[Section 7.2,][]{2013ApJS..208....4P}.
To model overshooting, we extend the hydrogen convective core boundary by 0.335 times the pressure scale height \citep{2011A&A...530A.115B}. 
\mesa\ uses the formulation of \citet{1983A&A...126..207L} to model semi-convection and \citet{1980A&A....91..175K} to model thermohaline mixing.
We set the value of $\alpha_{\rm sc}$ to 1.0 \citep{2006A&A...460..199Y} and $\alpha_{\rm th}$ to 1.0, respectively.

Stellar evolution codes use theoretical and empirical prescriptions to determine wind mass-loss rates using properties of the stellar photosphere. We use a customized `Dutch' wind mass-loss prescription in \mesa\footnote{The `Dutch' wind mass-loss scheme is a combination of the prescriptions of \citet{2001A&A...369..574V}
(when $T_{\rm eff} \gtrsim 10^4\,$K and $X_{\rm surf}$ $\geq$ 0.4), \citet{2000A&A...360..227N} (when $T_{\rm eff} \gtrsim 10^4\,$K and $X_{\rm surf}$ $<$ 0.4),
and \citet{1988A&AS...72..259D} (when $T_{\rm eff} \leq 10^4\,$K)}, by scaling the prescription of \citet{1988A&AS...72..259D} by $(Z/Z_\sun)^{0.85}$ to match the metallicity scaling of \citet{2001A&A...369..574V}, where $Z$ is the initial metallicity of the model. 
As discussed in paper I, some studies suggest that red supergiant winds are independent of metallicity \citep{2000A&A...354..125V,2005A&A...438..273V,2017MNRAS.465..403G}. 
Using stronger cool winds at low metallicity would cause single stars to resemble their solar metallicity counterparts, without affecting binary SNe IIb, as their evolution is dominated by mass transfer and stronger Wolf-Rayet (WR) winds (see paper I). 
We assume that stellar winds carry away the specific orbital angular momentum of the mass losing binary component.

In binaries, we compute mass loss due to Roche-lobe overflow (RLO) using the prescription of \citet{1990A&A...236..385K}. 
A fixed fraction, $\epsilon$, of the mass lost from the primary due to RLO is transferred to the secondary and the rest is assumed to be lost to the CSM as the secondary's stellar wind. 
In this paper, we focus on binary models with $\epsilon=$ 0.5 and 0.1, though we also have additional models with $\epsilon$ = 1.0 and 0.01 at solar metallicity (see paper I).
We ignore models with $\epsilon$ = 1.0 because mass transfer is expected to be non-conservative as a result of the spin-up of the secondary to critical rotation due to accretion \citep[][also see \citet{1991ApJ...370..604P} for a counter argument]{1981A&A...102...17P,2005A&A...435.1013P,2012A&A...537A..29R}.
Moreover, in paper I we show that binary SNe IIb should have low mass transfer efficiencies in order to explain observed SN IIb fractions.
Finally, we ignore models with $\epsilon$ = 0.01 because, as we show in paper I, they occupy roughly the same parameter space as those with $\epsilon$ = 0.1.
All orbits are circular.

We track the evolution of all our SN IIb models from zero-age main sequence (ZAMS) to core carbon exhaustion. 
We define SN IIb progenitors as those that reach core carbon exhaustion with $0.01-1M_\sun$ of residual hydrogen envelope\footnote{The hydrogen envelope boundary is defined as the outermost Lagrangian co-ordinate where the hydrogen mass fraction $\leq 0.01$ and the helium mass fraction $\geq 0.1$.}. This definition is a conservative choice. Inferred hydrogen envelope masses for SNe IIb, including those with detected progenitors, are $\lesssim 0.5 M_\odot$ (see Table \ref{t:SNIIbprops}). Since we are observationally biased towards finding the progenitors of SNe IIb with more massive envelopes \citep[as they are more extended][and redder]{2017ApJ...840...10Y}, they represent progenitors with the most massive envelopes. 
Also, while detailed radiative transfer calculations show that progenitors with total hydrogen mass $\gtrsim0.001M_\sun$ yield the characteristic SN IIb spectra at early times \citep{2011MNRAS.414.2985D}, there are many SNe Ib with evidence of high-velocity hydrogen during early phases [e.g., SN 2008D \citep{2008Sci...321.1185M,2009ApJ...702..226M} and SN 2014C \citep{2015ApJ...815..120M,2017ApJ...835..140M}].
It is also important to remember that envelope mass constraints from LC modeling are not robust (see also Section \ref{s:obs}).
We note that results in this paper are not sensitive to alternative more restrictive definitions considered in paper I.
Though we do not track the evolution of secondaries after primary CC, some secondaries can produce SNe IIb. 
We expect these to resemble single SN IIb, because 80\% of binaries are expected to become unbound after primary explosion \citep{2019A&A...624A..66R}.
In systems that do remain bound, RLO is expected to result in unstable mass transfer and merger due to the extreme mass ratio between the secondary and the compact object remnant.

We classify binary SNe IIb into three types based on their interaction history. 
Case early-B (EB) SNe IIb initiate mass transfer while the primary is crossing the Hertzsprung Gap, case late-B (LB) SNe IIb initiate mass transfer after the primary begins its rise on the giant branch but before it exhausts helium in its core, and case C SNe IIb undergo mass transfer after the primary exhausts helium in its core.

We make the following assumptions regarding the birth properties of single and binary stars in the Universe.
We assume a Salpeter IMF \citep{1955ApJ...121..161S}
\begin{equation} \label{e:IMF}
f(M_{\rm ZAMS}) = M^{-\alpha} .
\end{equation}
with $\alpha =2.3$ \citep{2001MNRAS.322..231K}.
We adopt a power-law distribution for the initial mass ratio, $q$, 
\begin{equation} \label{e:f(q)}
f(q_{\rm ZAMS}) = q^\beta .
\end{equation}
with $\beta= 0.0$ \citep{2012Sci...337..444S, 2014ApJS..213...34K}.
We assume a power-law distribution for the initial orbital period, $P_{\rm orb}$,
\begin{equation} \label{e:f(P)}
f(\logten P_{\rm orb})=(\logten P_{\rm orb})^{\gamma} .
\end{equation}
where $\gamma=-0.22$ \citep{2014ApJS..213...34K}.
We do not make an assumption regarding binary fraction since we compare single and binary SN IIb populations to observed constraints separately.

\section{Correlations Between Progenitor Properties} \label{s:relations}

\begin{figure}
\begin{center}
\includegraphics[width=\columnwidth]{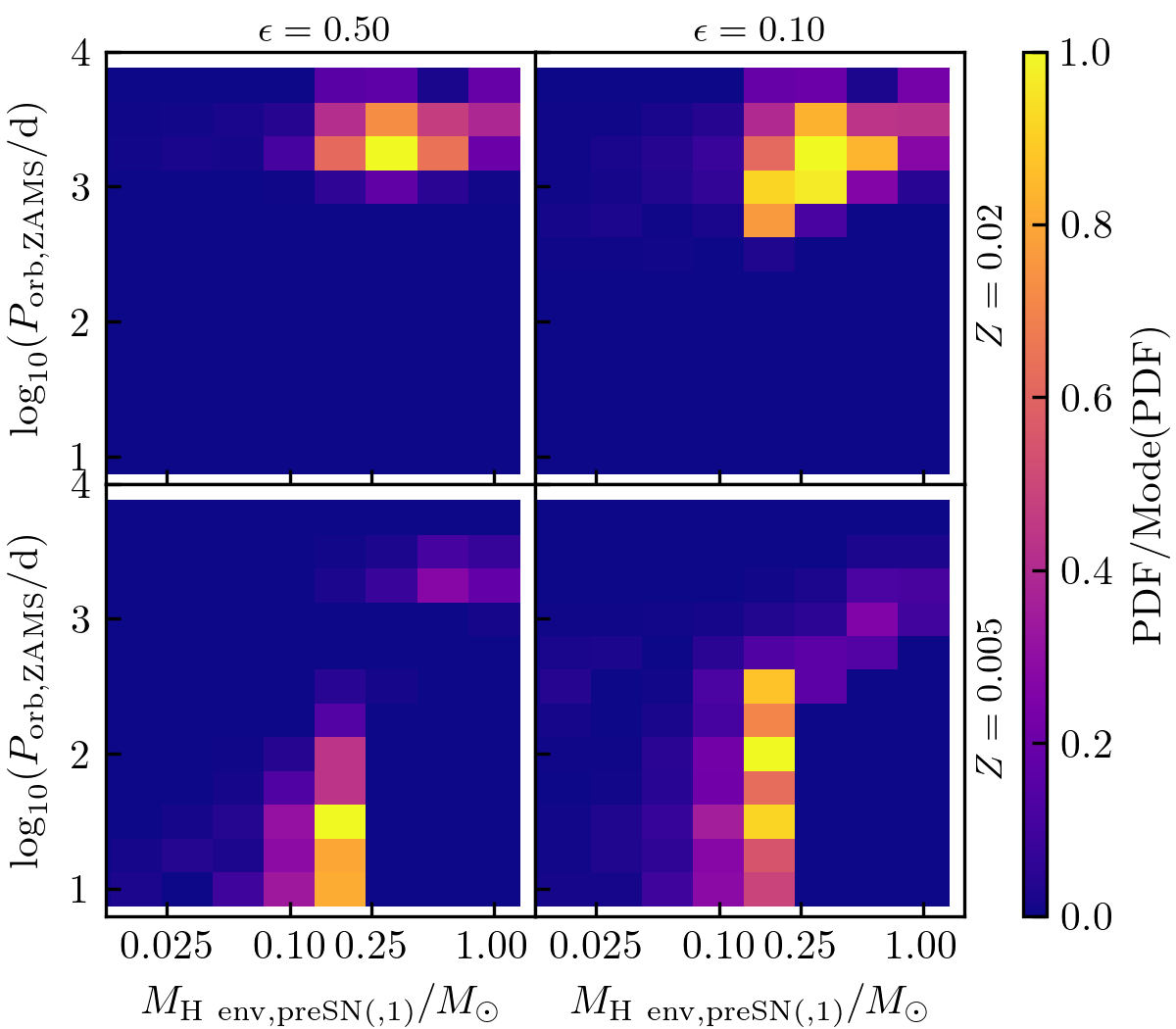}
\end{center}
\caption{Distribution of initial orbital period ($P_{\rm orb, ZAMS}$) as a function of pre-SN progenitor hydrogen envelope mass ($M_{\rm H~env,preSN(,1)}$) for binary SNe IIb with $\epsilon$ = 0.5 (left) and 0.1 (right) at solar (top) and low metallicity (bottom).
\label{f:P-env}}
\end{figure}

Figure \ref{f:P-env} shows the probability distribution of initial orbital period as a function of pre-SN progenitor hydrogen envelope mass for binary SNe IIb at solar and low metallicity.
Binaries with shorter initial orbital periods result in more stripped SN IIb progenitors. 
This relation has been previously discussed by \citet{2017ApJ...840...10Y} and \citet{2017ApJ...840...90O}.
At solar metallicity, SNe IIb typically have larger envelope masses and a wider range in values.
The biggest difference between SN IIb populations at solar and sub-solar metallicities arises due to case EB binaries.
As we show in paper I, while solar metallicity EB binaries are completely stripped before explosion, their low metallicity counterparts are able to evolve to CC without getting stripped, owing to weaker winds. 
The low-metallicity EB SNe IIb are highly stripped with small residual envelope masses $\sim 0.1-0.2M_\odot$ (see also discussion about SN IIb definitions in Section \ref{s:models}).
EB SNe IIb dominate the distribution at low metallicity due to the large range in initial orbital periods that permit mass transfer during this phase.
However, it is likely that this group also exists at solar metallicity given evidence that winds are weaker than previously assumed \citep[][see also discussion in Section 6.2 in paper I]{2014ARA&A..52..487S, 2017ApJ...840...10Y}.
In fact, as we note in paper I, the existence of case EB SNe IIb at solar metallicity is needed to account for observed SN IIb fractions.


\begin{figure}
\begin{center}
\includegraphics[width=\columnwidth]{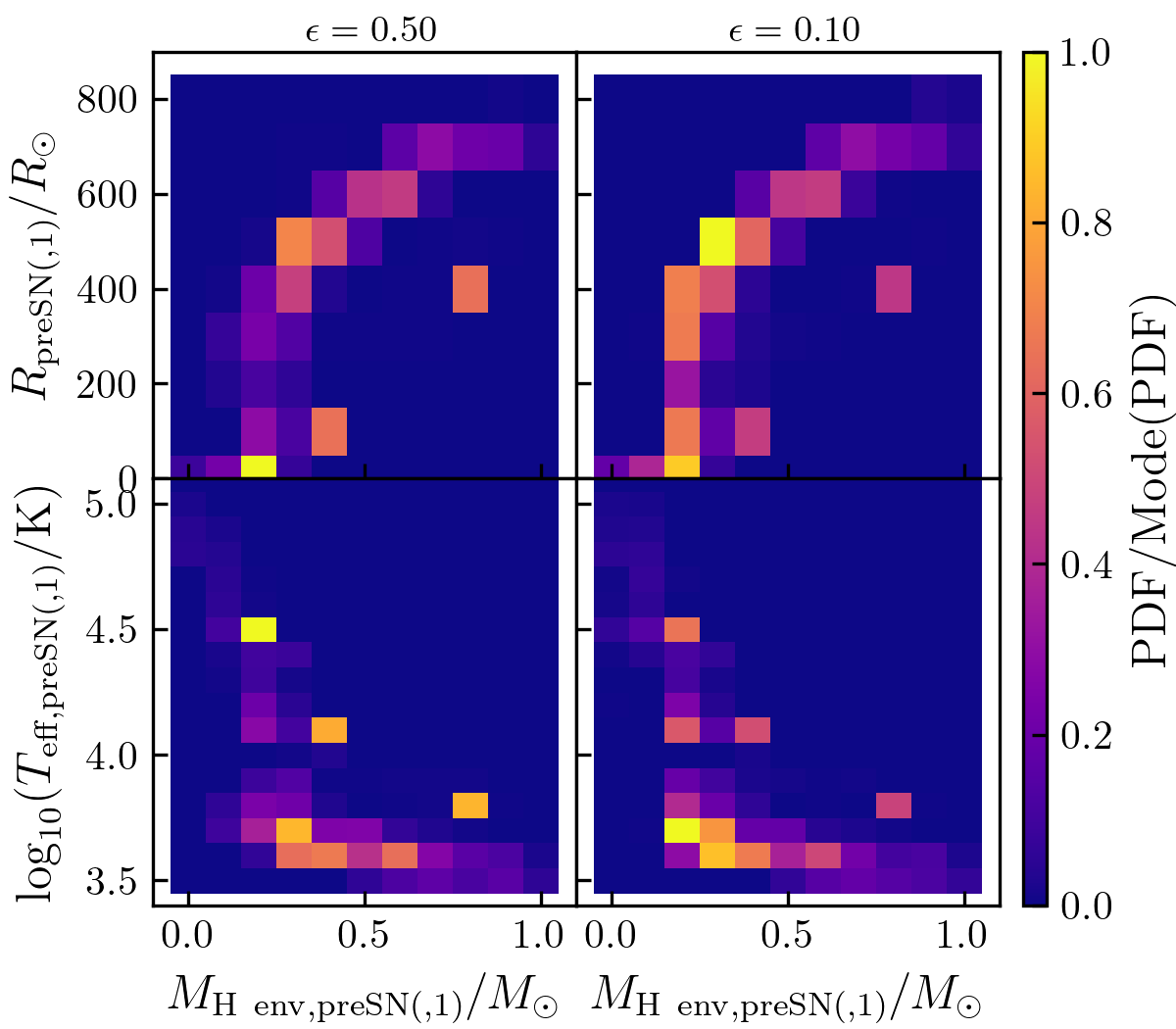}
\end{center}
\caption{Distribution of pre-SN progenitor radius ($R_{\rm preSN(,1)}$, top) and effective temperature ($T_{\rm eff, preSN(,1)}$, bottom) as a function of pre-SN progenitor hydrogen envelope mass $M_{\rm H~env,preSN(,1)}$, left) for binary SNe IIb with $\epsilon$ = 0.5 (left) and 0.1 (right) at solar metallicity.
SNe IIb progenitor with smaller (larger) envelope masses are more compact (extended) and hot (cool).
\label{f:R-T-env}}
\end{figure}

Figure \ref{f:R-T-env} shows the probability distribution of pre-SN progenitor radius and effective temperature as a function of its hydrogen envelope mass for solar metallicity binary SNe IIb.
SNe IIb progenitors with smaller (larger) envelope masses are more compact (extended) and hot (cool). 
The phenomenon can be understood as more progressive stripping causing SN IIb progenitors to increasingly resemble hydrogen-deficient WR stars.
This correlation was previously discussed by \citet{2015A&A...575A..60M} and \citet{2017ApJ...840...10Y}.
However, the correlation is less apparent in the population of low metallicity SNe IIb, once again due to the dominant group of EB SNe IIb. 
We do not show the corresponding Figure \ref{f:R-T-env} for low metallicity SNe IIb because it contains a single strongly peaked distribution at small envelope masses and radii (see lower left and right panels, respectively, of Figure \ref{f:env-core-R} for reference). 
However, as we mention earlier, we expect that this distribution is followed even at solar metallicity

\begin{figure}
\begin{center}
\includegraphics[width=\columnwidth]{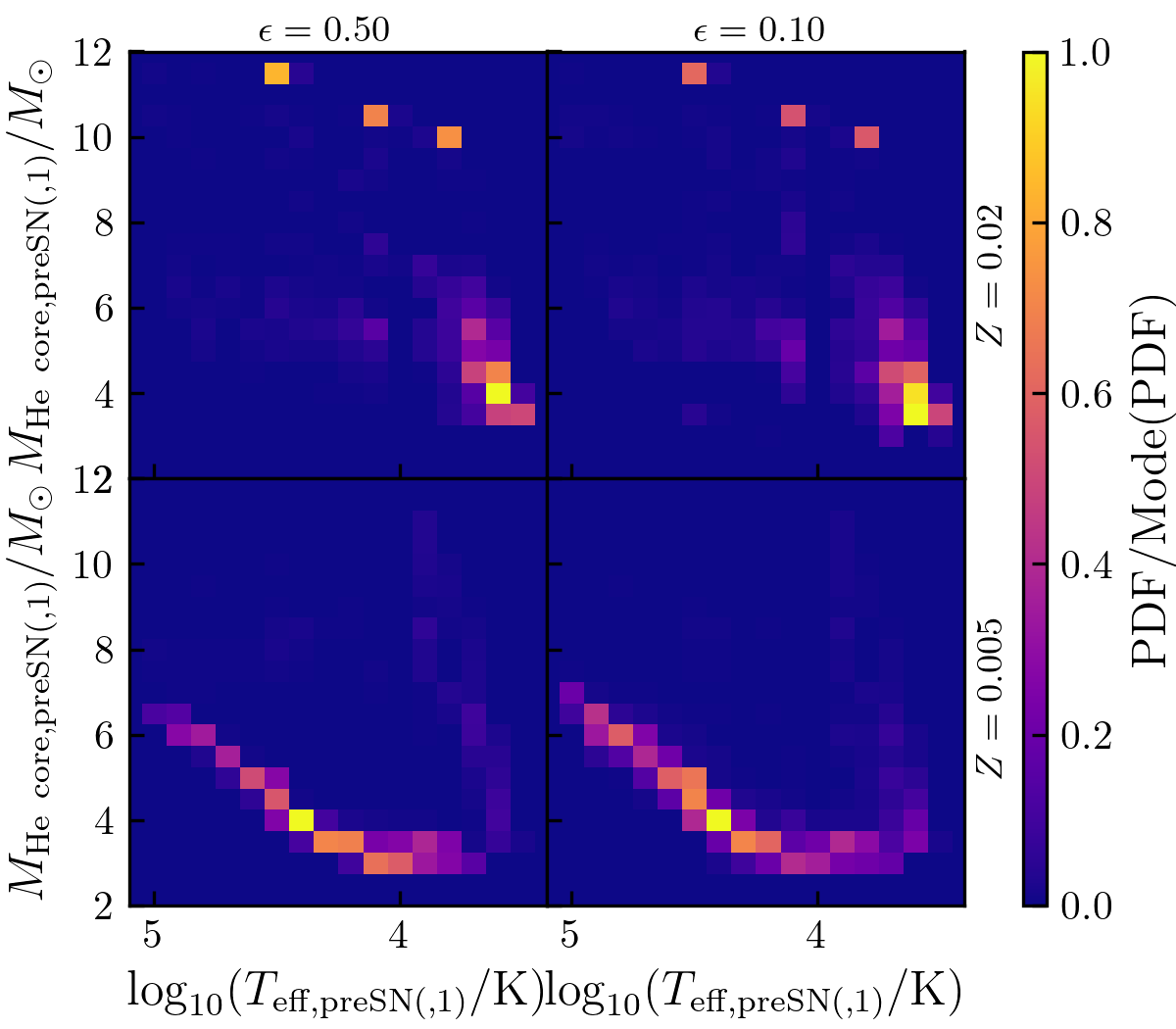}
\end{center}
\caption{Distribution of pre-SN progenitor helium core mass ($M_{\rm He~core,preSN(,1)}$) as a function of its effective temperature ($T_{\rm eff, preSN(,1)}$) for binary SNe IIb with $\epsilon$ = 0.5 (left) and 0.1 (right) at solar (top) and low metallicity (bottom).
\label{f:core-Teff}}
\end{figure}

Figure \ref{f:core-Teff} shows the probability distribution of pre-SN progenitor helium core mass as a function of its effective temperature for binary SNe IIb at solar and low metallicity. 
The correlations in this space closely follow the Hertzsprung-Russell (H-R) diagram (discussed in the next section, see also Figure \ref{f:HR}) since the SN IIb progenitor luminosity is determined by its helium core mass. 
The range in helium core masses is roughly similar at both metallicities as these are formed during the main sequence and our binaries experience mass transfer after core hydrogen exhaustion.
While solar metallicity SNe IIb are mostly cool and extended, low metallicity SNe IIb are hot and compact.
As before, the distribution of low metallicity SNe IIb are dominated by EB SNe IIb. They produce a tight correlation in this space due to the helium main sequence (discussed more in Section \ref{s:results}). 

\section{Observational Constraints for SNe II\lowercase{b}} \label{s:obs}

Table \ref{t:SNIIbprops} lists observationally constrained properties for various SN IIb progenitors using three observational probes: from multi-band LCs, direct progenitor detections, and X-ray/radio observations.
Before comparing our models to these constraints, we discuss 
the methodology involved in extracting constraints from each of these probes to identify potential sources of bias and uncertainty relevant to our analysis.

\subsection{Constraints from LCs}

The most widely used observational probe into SN progenitors is analyzing multi-band LCs. 
Most SNe IIb exhibit a characteristic two-peak optical LC. The first peak is due to the SN shock breaking out of the stellar surface while the second is due to radioactive heating from the decay of $^{56}$Ni and $^{56}$Co \citep{1993Natur.364..507N,1993Natur.364..509P,1994A&A...281L..53B,1994A&A...281L..89U,1994ApJ...429..300W}.
The first peak is not always observed, as in the cases of SNe 2008ax and 2015as, if the envelope size is too small leading to a short lived shock breakout phase.
SN IIb LCs can be modeled either (semi-)analytically or via hydrodynamical SN simulations.

The simplest approach for constraining SN IIb progenitor properties is fitting the radioactive decay powered phase of its LC using the model of \citet{1982ApJ...253..785A}.
Analytical SN models of \citet{2014ApJ...788..193N} and \citet{2015ApJ...808L..51P} 
are tailored to SNe IIb and connect features of their LCs (e.g. time and luminosity of the first peak) to the progenitor structure (e.g. envelope mass and radius). 
Semi-analytic models of \citet{2016A&A...589A..53N} and \citet{2017ApJ...838..130S} use numerical calculations to improve approximations used in analytic models. 
Radiative hydrodynamical SN models involve injecting energy near the center of stellar progenitor models, adding energy from radioactive decay of Ni and following the ensuing shock-wave as it propagates through the stellar structure, to shock break-out and later nebular phases. 

These models constrain the progenitor envelope and core mass and radius.
It has been observed that progenitor constraints derived from analytic and semi-analytic models typically yield smaller values than those from hydrodynamical modeling \citep[][but see \citep{2018A&A...609A.136T}]{2009A&A...506..829U,2009MNRAS.395.1409S,2017ApJ...837L...2A}.
Moreover, constraints derived from analytic and semi-analytic models are strongly dependent on assumed parameters (e.g. opacity) and may not always agree with one another.
Due to these differences, in Table \ref{t:SNIIbprops} constraints derived using analytic/semi-analytic models are noted.

While Table \ref{t:SNIIbprops} lists constraints derived for individual SNe, recently, given the increase in the sample of SN LCs due to surveys, there have been efforts to study LCs for a set of SE SNe \citep{2016MNRAS.457..328L,2018A&A...609A.136T,2019MNRAS.485.1559P}.
These studies use the model of \citet{1982ApJ...253..785A} to constrain the SN ejecta mass\footnote{\citet{2018A&A...609A.136T} also used hydrodynamical models and found similar results to those from using the simple Arnett prescription.}.
Assuming a 1.5 $M_\sun$ NS remnant, the ejecta mass can then be used to get a rough estimate for the progenitor core mass. 
Though approximate, constraints from these studies have the advantage of being homogeneously reduced (e.g. when constructing bolometric LCs).
We consider these constraints separately when comparing our models to observations.

\subsection{Constraints from Progenitor Identifications}

Another, albeit serendipitous, probe into SN progenitors is direct detections in archival images.
A progenitor detection in more than one filter can constrain the progenitor luminosity and effective temperature. Direct progenitor detections thus provide an independent probe of the progenitor radius from analyzing LCs.

SNe IIb have a higher number of progenitors detected in pre-SN images compared to other SE SN types. To date, five SN IIb progenitor candidates have been identified in pre-explosion images [SNe 1993J \citep{1994AJ....107..662A, 2004Natur.427..129M,2014ApJ...790...17F}, 2008ax \citep{2008MNRAS.391L...5C,2015ApJ...811..147F}, 2011dh \citep{2011ApJ...739L..37M,2011ApJ...741L..28V,2013ApJ...762...74B}, 2013df \citep{2014AJ....147...37V,2015ApJ...807...35M}, and 2016gkg \citep{2017MNRAS.465.4650K,2017ApJ...836L..12T,2018Natur.554..497B}], compared to one each for the other two SE SN types: Type Ib SN iPTF13bvn \citep{2013ApJ...775L...7C, 2016ApJ...825L..22F} and Type Ic SN 2017ein \citep{2018arXiv180301050V, 2018MNRAS.480.2072K}. In addition, there is evidence for the presence of binary companions to the progenitors of SNe 1993J \citep{2014ApJ...790...17F}, 2001ig \citep{2018ApJ...856...83R}, and 2011dh \citep{2014ApJ...793L..22F}.

While powerful, progenitor constraints derived from direct detections have important limitations. 
The primary uncertainty is due to line-of-sight coincidences and these can be difficult to quantify.  
It is also difficult to rule out flux contamination from nearby stars and/or a binary companion. 
Another source of uncertainty comes from using stellar atmosphere models to constrain the progenitor SED. 
The properties of the model stellar atmospheres are strongly dependent on their mass loss rates \citep{2018A&A...615A..78G}. 
However, wind mass-loss rates, especially during late evolutionary phases, are highly uncertain \citep{2014ARA&A..52..487S}. 
There is also added uncertainty from potential mass loss due to binary RLO. 
Finally, since most progenitors are discovered in high-resolution HST archival images, there is selection bias against hot progenitors since the instrument lacks filters in wavelengths shorter than $\sim 220$ nm. 

\subsection{Constraints from X-rays/Radio}

\begin{figure*}
\begin{center}
\includegraphics[width=\textwidth]{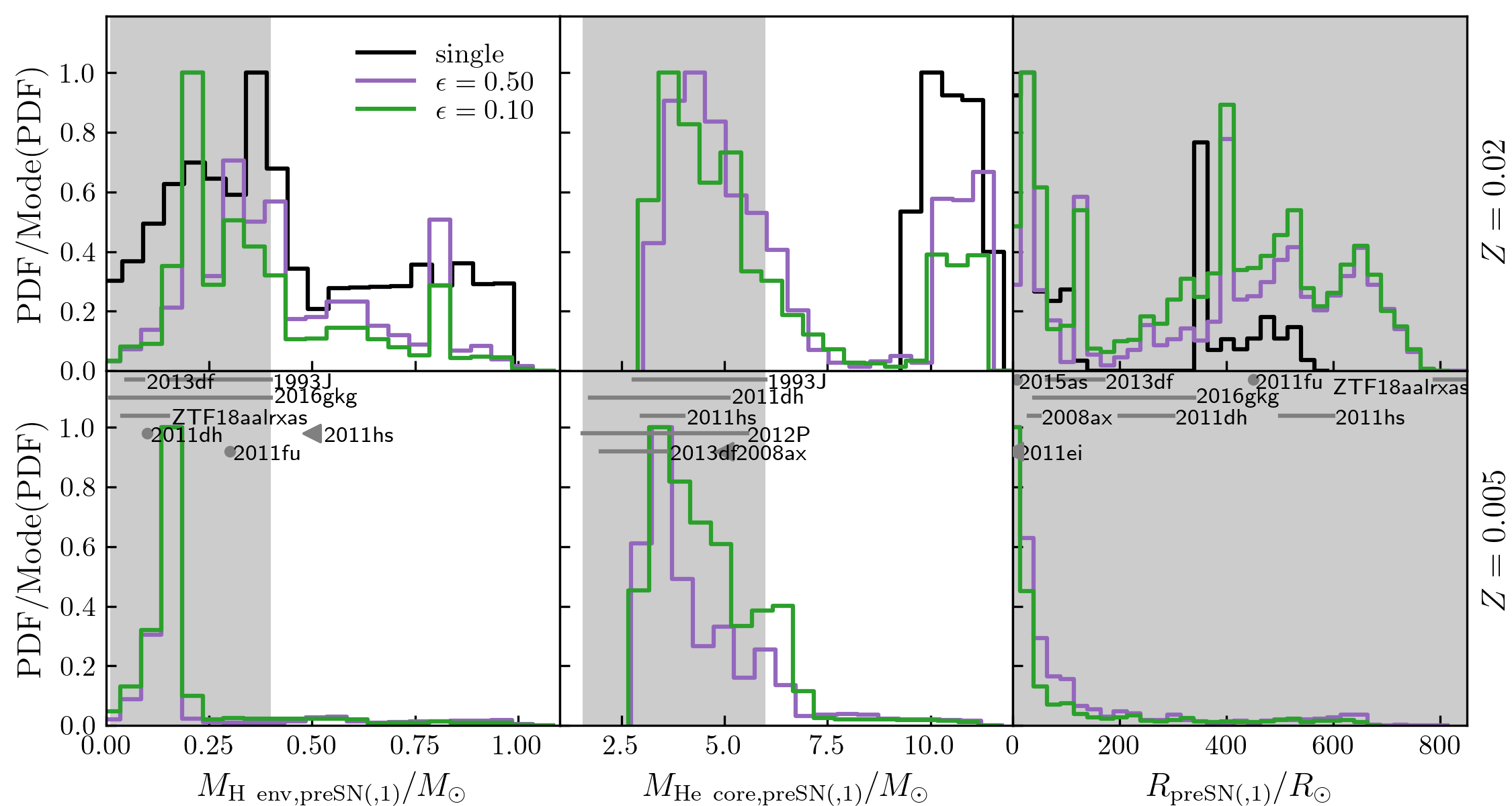}
\end{center}
\caption{Distributions of pre-SN progenitor hydrogen envelope mass ($M_{\rm H~env,preSN(,1)}$, left), helium core mass ($M_{\rm He~core,preSN(,1)}$, middle), and radius ($R_{\rm preSN(,1)}$, right) for single (black) and binary [$\epsilon$ = 0.1 (green) and 0.5 (purple)] SN IIb progenitors at solar (top) and low metallicity (bottom).
Grey shaded regions indicate the range of values derived for SNe IIb listed in Table \ref{t:SNIIbprops}. 
Constraints for individual events are indicated using lines (ranges), circles (single values), and triangles (upper limits).
\label{f:env-core-R}}
\end{figure*}

\begin{figure*}
\begin{center}
\includegraphics[width=\textwidth]{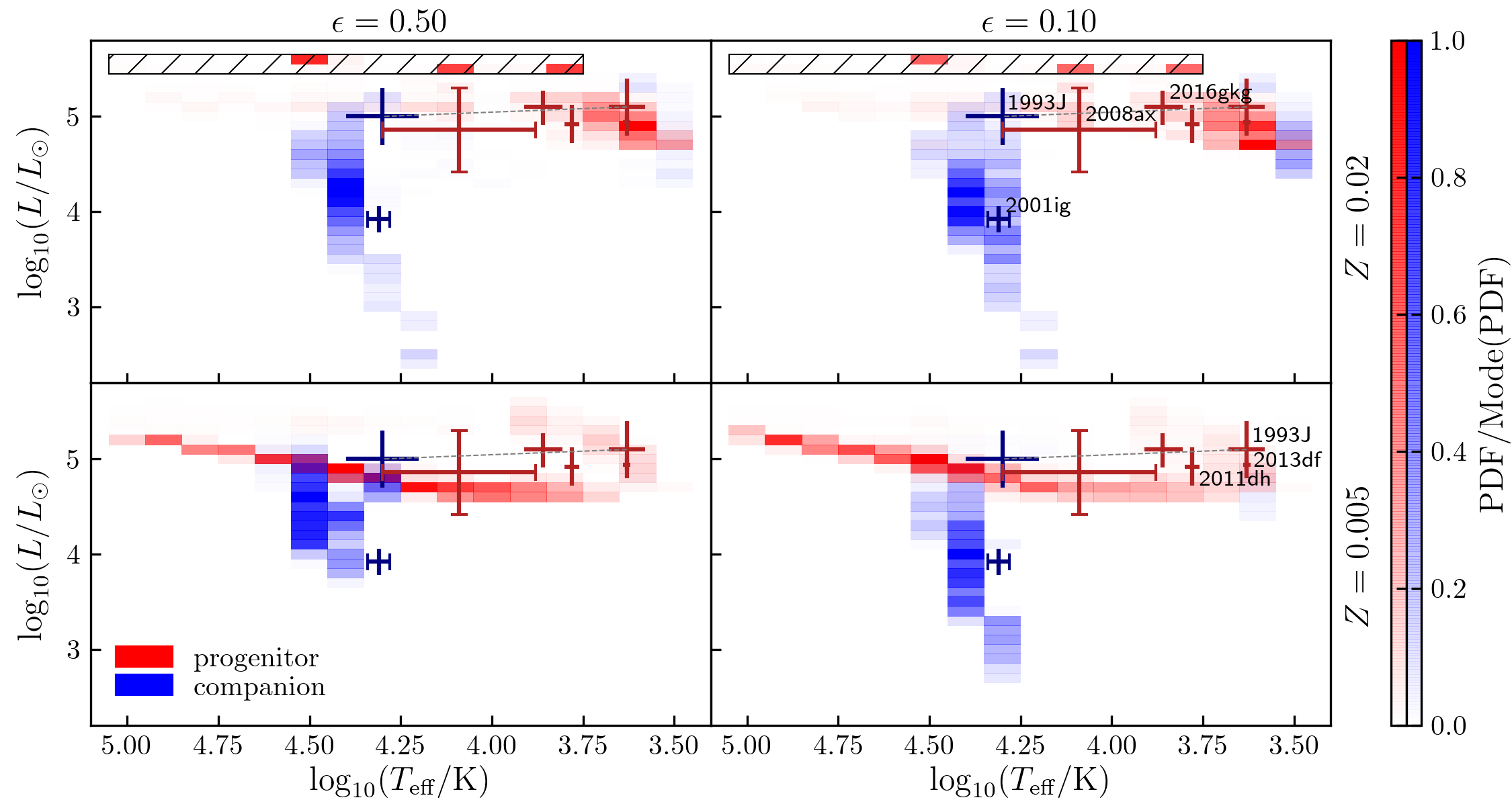}
\end{center}
\caption{Distributions of pre-SN H-R locations of single and binary ($\epsilon$ = 0.5 (left) and 0.1 (right)) SN IIb progenitors at solar (top) and low metallicity (bottom). 
The red (blue) color scale shows probability for binary SN IIb progenitors (companions). 
The black hatched region shows the space spanned by single SN IIb progenitors with probability distribution divided by its mode $\geq 0.01$.
Red cross-hairs show H-R locations of SN IIb with progenitor identifications. 
Blue cross-hairs show inferred H-R locations of the companion of SNe 1993J and 2008ax.
Cross-hairs without caps indicate 1$\sigma$ uncertainties while those with caps indicate ranges.
The two groups in the progenitor H-R space at low metallicity, the first spanning across the figure and the second clustered at the top-right, are due to case EB and case LB/C SNe IIb, respectively.
\label{f:HR}}
\end{figure*}

The third observational probe into SN progenitor properties is from X-ray/radio observations.
Non-thermal X-ray and radio emission arises as the SN shock interacts with the circumstellar medium (CSM) shaped by the progenitor.
X-ray emission arises when CSM electrons are shocked to relativistic speeds, either by the forward or the reverse shock, due to inverse Compton, Bremsstrahlung, or synchroton processes \citep[e.g.][]{1996ApJ...461..993F,2016ApJ...818..111K}. 
Both the forward and reverse shocks can contribute to emission in X-rays. 
Radio emission arises when shock accelerated electrons gyrate around shock amplified magnetic fields producing synchrotron emission.
X-ray and radio emission provide independent constraints on the CSM density if they arise from different processes (e.g. if radio is due to synchrotron and X-rays are either due to inverse Compton or Bremsstrahlung).

Once again there are important limitations to consider when interpreting constraints derived from X-ray/radio analyses.
First, observations help constrain CSM densities at the radius from which the emission arises\footnote{However, if the emission arises from free-free absorption in radio or photoelectric absorption in X-rays they probe the material in front of the shock.}, $\rho(R)$. While a progenitor mass-loss rate can be derived assuming a wind-like CSM profile as $\dot{M}_{\rm wind}=4\pi R^2 \rho(R) v_{\rm wind}$, these can have significant uncertainties.
For example, the progenitor wind velocity can vary by two orders of magnitude depending on whether the progenitor was a red supergiant or a WR star.
Moreover, the assumption of wind-like profile itself may not be accurate if there are density fluctuations resulting from intermittent mass loss episodes, either due to inefficient mass loss in binary interactions and/or stellar eruptions.
In other words, the treatment washes out information on the actual density profile probed by the observations. 
In fact, \citet{2013ApJ...767...71M} found evidence of CSM density modulations by a factor of two for SN 2011ei.
\citet{2014Natur.509..471G} found evidence of eruptive mass loss in the immediate environment ($\sim10^{12}$cm) of the progenitor of SN 2013cu using flash spectroscopy.
Note however that \citet{2015ApJ...811..117S} did not find evidence for pre-SN eruptions for the majority of SNe IIb.
Therefore, in this work, we use the ratio of $\dot{M}_{\rm wind}$ and $v_{\rm wind}$, as this is the only reliable observationally derived quantity.
Second, emission from both the near and far side of the SN shock, potentially arising from different environments, can contribute during later phases.
Therefore, constraints from earlier in the SN evolution are cleaner probes of CSM density.
All values listed in Table \ref{t:SNIIbprops} are derived for CSM at radii $\lesssim10^{16}$cm. 
Finally, it is important to remember that CSM density estimates are strongly dependent on assumed shock microphysics,  which are typically loosely constrained.

\section{Distributions of SN II\lowercase{b} Progenitor Properties and Comparison to Observational Constraints} \label{s:results}


We compare the probability distributions of progenitor properties to observationally constrained values from all three probes discussed in Section \ref{s:obs}.
Given the inconsistencies in statistical characteristics of derived values in Table \ref{t:SNIIbprops}, unless otherwise noted, we convert all constraints with uncertainties to a `range' using the 3$\sigma$ spread. 
This choice was intended to keep our inferences conservative.
We do not use the CSM density constraint for SN 2008ax as its $3\sigma$ lower limit is 0.
We note that the distributions are not sensitive to alternative more restrictive definitions considered in paper I.

\newpage
\subsection{Properties from Light Curves} \label{ss:obsLC}

Figure \ref{f:env-core-R} shows distributions of single and binary SN IIb pre-SN progenitor properties at solar and low metallicity that can be constrained using LCs: hydrogen envelope mass, helium core mass, and radius. 
The distribution of helium core mass for binary SN IIb progenitors at solar metallicity is bimodal. 
The high (low) helium core mass peak is due to progenitors with $M_{\rm ZAMS,1} \gtrsim (\lesssim) 20 M_\sun$. 
The high core mass group consists of mildly interacting binaries whose evolution largely resembles their single-star counterparts.
This group does not exist at low metallicity because progenitors with $M_{\rm ZAMS,1} \gtrsim 20 M_\sun$ do not evolve to SNe IIb (see paper I).
The low helium core mass peak is similar at both metallicities as our binaries experience mass transfer after the main sequence.

Observational constraints for helium core masses of SNe IIb are $1.5-6 M_\sun$. 
This constraint excludes single stars and the binary high core mass group at solar metallicity as SNe IIb progenitors. 
This conclusion has been underscored extensively in the literature as one of the key pieces of evidence supporting binaries as SN IIb progenitors. 
We also consider constraints derived from samples of SNe IIb \citep{2016MNRAS.457..328L,2018A&A...609A.136T,2019MNRAS.485.1559P}. We do not directly compare to derived distributions because these are not corrected for sampling bias or completeness and thus cannot be directly compared to SNe IIb populations. 
The 3$\sigma$ range in helium core masses derived from these studies is $1.5-11.8 M_\sun$ (assuming a NS remnant of mass $1.5M_\sun$). 
This range is broad and does not constrain our models.
Interestingly, \citet{2019MNRAS.485.1559P} found evidence of two peaks in the distribution of SN IIb ejecta masses, at 2 and 4$M_\sun$ (corresponding to 3.5 and $5.5M_\sun$ for helium core masses). 
While our distribution is also bimodal, they peak at 3.5 and $11M_\sun$. 
As we highlight in Section \ref{s:obs}, since \citet{2019MNRAS.485.1559P} used the analytic model of \citet{1982ApJ...253..785A}
the derived values are likely only accurate within a factor of a few. 

The distribution of progenitor radii of binary SN IIb progenitors at solar metallicity is also bimodal. 
The low (high) radius peak is due to progenitors with $T_{\rm eff} > (<) 10^4$ K.
The peak in the radius distribution around $400 R_\sun$ for high metallicity SNe IIb is due to the Hayashi line allowing progenitors with increasing luminosity/decreasing envelope mass to be stable with the same radius.
Binary SN IIb progenitors at low metallicity have smaller hydrogen envelope masses and radii (a peak in progenitor envelope mass translates to a peak in radius, and vice versa; see Section \ref{s:relations}) than at solar metallicity. 
This is due to the group of EB SNe IIb. 
As discussed earlier, this channel is only viable in low metallicity models where weak winds permit the progenitor to retain its hydrogen envelope left over after the mass transfer phase.
The group of EB SNe IIb dominate the distribution at low metallicity because the parameter space that permits case EB mass transfer is large (see paper I).
We note that it is likely that low metallicity models also represent solar metallicity environments.
There is evidence that stellar wind prescriptions, including those used in this work, are overestimated \citep{2014ARA&A..52..487S,2019A&A...632A.126S,2020arXiv200806066B} particularly for late-stage low-mass stars \citep[][see also discussion in Section 6.2 in paper I]{2017ApJ...840...10Y}.
Moreover, as suggested in paper I, low metallicity models can help account for observed SN IIb fractions irrespective of metallicity.

Observational constraints exclude SNe IIb with more massive envelopes at solar metallicity. 
However, as we note in Section \ref{s:models}, excluding models with hydrogen envelope $>0.5M_\sun$ do not change the overall distributions presented in this paper much.
Observations indicate a broad range for SN IIb progenitor radii and as such do not provide strong constraints for our models.
Although compact progenitors are not represented in observations, this could be a result of observational bias (see Section \ref{s:obs}). 
Note also that SNe IIb from compact progenitors would be harder to detect because the cooling envelope feature in their LCs would be less pronounced \citep{2016MNRAS.455..423M}.
The existence of compact progenitors is favored by our models and, if present, especially in high metallicity environments, would indicate the presence of weak stellar winds.

\subsection{Properties from Detection of Progenitors and their Companions}

Figure \ref{f:HR} shows distributions in the H-R diagram of single and binary SN IIb progenitors at solar and low metallicity.
There are two groups in the binary SN IIb progenitor H-R space at low metallicity. 
The top-right group is roughly similar to the one at solar metallicity and is due to case LB/C SNe IIb. 
The bottom-left group, spanning across the H-R diagram, is due to case EB SNe IIb. 
Since these progenitors are highly stripped they lie along the helium MS \citep{2015A&A...573A..71K}.
Although, these progenitors are not represented in observations, this could be due to observational bias towards detecting redder stars (see Section \ref{s:obs}).
Note that there is indirect evidence for a WR progenitor for SN 2013cu using flash spectroscopy \citep[][but see \citet{2014A&A...572L..11G}]{2014Natur.509..471G}. 
There is a gap in the H-R locations of the two groups at $\epsilon=0.5$. This is because progenitors that initiate mass transfer late on the HG with high mass transfer efficiency enter contact and are expected to merge (see paper I).

\begin{figure*}
\begin{center}
\includegraphics[width=\textwidth]{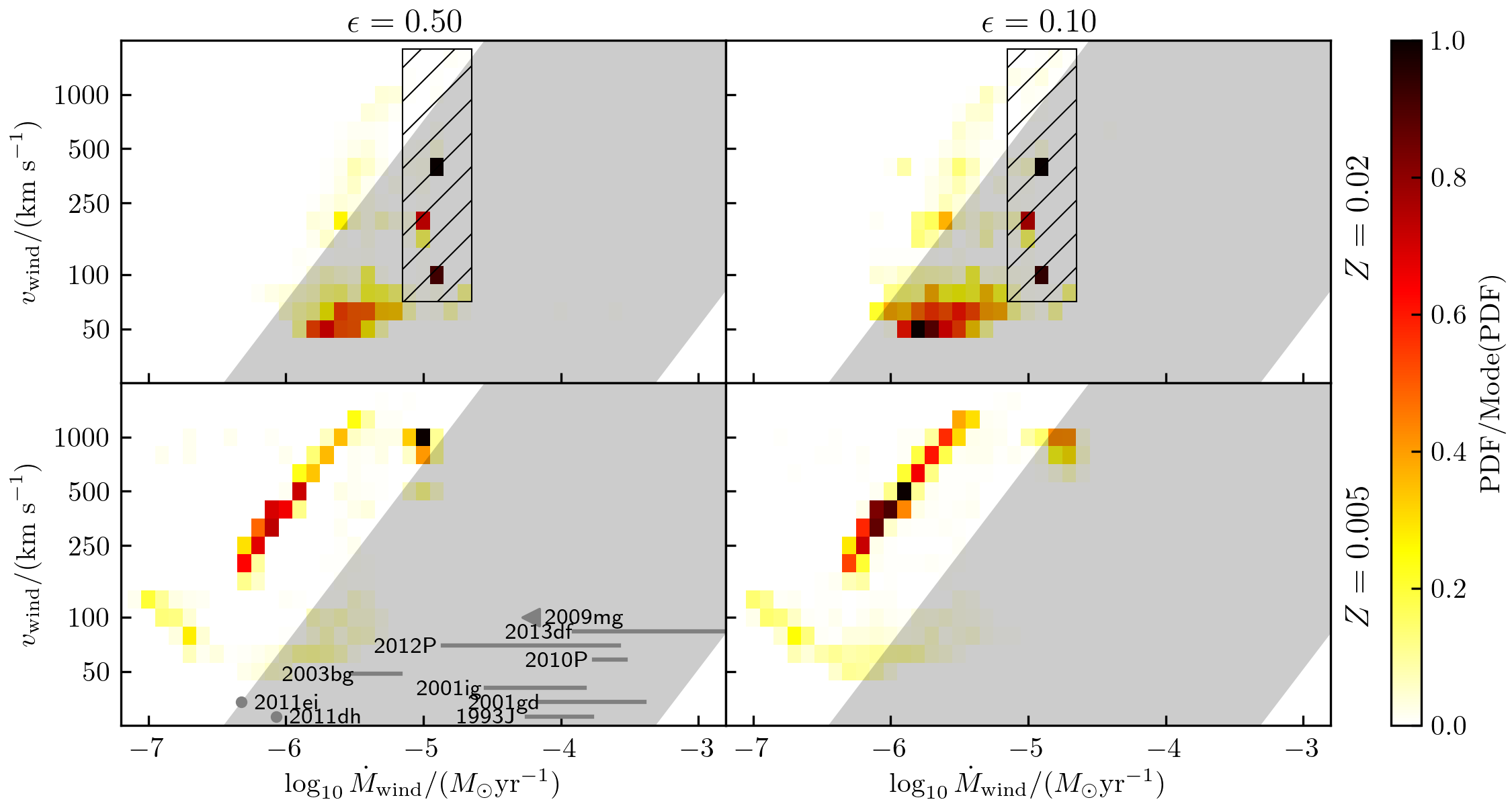}
\end{center}
\caption{Distributions of CSM properties of single and binary ($\epsilon$ = 0.5 (left) and 0.1 (right)) SNe IIb at solar (top) and low metallicity (bottom) shaped by the progenitor system 100 years before progenitor CC.
Black hatched regions show region occupied by single SNe IIb with probability distribution divided by its mode $\geq 0.01$.
Grey shaded regions show the observationally range inferred for SNe IIb. 
Constraints for individual events are indicated using lines (ranges), circles (single values), and triangles (upper limits).
Note that, as discussed in Section \ref{s:obs}, observations constrain CSM densities or the ratio of $\dot{M}_{\rm wind}$ and $v_{\rm wind}$.
\label{f:csm}}
\end{figure*}

Most companions lie on the main sequence, producing the almost vertical streak in the distributions. 
Similar to \citet{2011A&A...528A.131C}, we find that SN IIb companion effective temperature and luminosity decrease for lower accretion efficiencies. 
The range in companion luminosities for binary SN IIb with $\epsilon$ = 0.5 at low metallicity is smaller. 
This is because low initial mass ratio EB SNe IIb enter unstable mass transfer or contact due to a combination of the effects of low mass ratio, high mass transfer efficiency, and mass transfer on the Hertzsprung gap (see paper I). 
Redder companions result from binaries with initial mass ratios close to unity, where the companion also evolves off the main sequence.
These are less likely at higher mass transfer efficiencies, which produces more contact systems that we expect to merge.
We find that systems where the companion is red as a result of a recent rapid mass transfer event are rare. 

Overall, our binary SN IIb models are able to explain the H-R properties of SN IIb progenitors and companions detected to date.
H-R locations of detected progenitors of SNe IIb are favored at solar metallicity and low mass transfer efficiency ($\epsilon=0.1$) at low metallicity.
Single SN IIb progenitors are excluded within 1$\sigma$ of observed constraints. 
Our models indicate the existence of a large population of blue highly stripped SN IIb progenitors at low metallicity. 
However, as mentioned earlier, it is likely that these models represent solar metallicity environments as well. 

\subsection{Properties of Circumstellar Medium}

To compare CSM properties at similar observational conditions, we examine SN IIb progenitor properties a given number of years before CC.
Since our models are stopped at core carbon exhaustion, and this point could be between 0.1-100 years before progenitor CC depending on mass, we run single star models with stellar winds turned off until silicon core exhaustion\footnote{Defined as the point when the central silicon mass fraction drops below $10^{-6}$. Our models with ZAMS masses $\lesssim 10.5M_\sun$ and carbon core masses $\lesssim 2M_\sun$ ignite neon off-center and do not reach silicon core exhaustion. Since our our conclusions are not strongly dependent on the progenitor age at which the CSM is shaped, we ignore these models when computing fits.} to compute the time remaining until CC for a given core mass.
Specifically, we compute linear fits to time between core carbon and silicon exhaustion as a function of carbon core mass.
We assume that CSM properties are dominated by more massive wind parcels and that the wind velocity is the escape velocity of the star/binary component it originates from.
For the majority of binaries the progenitor star has the higher mass-loss rate, so its mass loss characterizes the CSM.

Figure \ref{f:csm} shows the distributions of CSM properties of single and binary SNe IIb at solar and low metallicity shaped by the progenitor system 100 years before progenitor CC.
Our results do not change much for CSM properties shaped either 10 or 1000 years before CC. 
CSM for binary SNe IIb at low metallicities is less dense than at solar. 
This is because binary SNe IIb progenitors at low metallicities are more compact and therefore have higher wind velocities. 
Case C SNe IIb are observationally valuable as they are more likely to be undergoing mass transfer close to CC, leaving an imprint of their interaction on their CSM. 
Although a few case C SNe IIb models with $\epsilon$ = 0.5 have their CSM affected due to binary interactions 1000 years before CC (with $v_{\rm wind}\sim1000$km s$^{-1}$ and $\dot{M}_{\rm wind}\sim10^{-4}M_\sun$ yr$^{-1}$), the overall distribution of CSM properties of case C SNe IIb are not very different from case B SNe IIb. 

Using four SN IIb with detected progenitors, \citet{2015ApJ...807...35M} found a correlation between CSM density and progenitor radius derived from photometry.  
\citet{2016ApJ...818..111K} found similar results using radio luminosity (which is a proxy for CSM density) and the product of the peak bolometric luminosity and duration of the shock breakout phase of the LC (which is proxy for progenitor radius).
We cannot verify these relations because our wind velocity estimate depends on the progenitor radius.
However, we note that, given the relation between SN IIb envelope mass and radius described in Section \ref{s:relations}, it is likely there is also an analogous correlation between CSM density and progenitor envelope mass.

Observationally constrained CSM properties for SNe IIb are consistent (though only on the low side) with single SN IIb progenitors at low metallicity and case LB/case C binary SN IIb progenitors at both metallicities.
However, since there is a bias toward detecting brighter events in X-rays/radio, the non-representation of the dominant group of EB SNe IIb could be due to observational bias. 
Interestingly, \citet{2020ApJ...891..116W} found evidence for a WR-like wind phase close to CC for the progenitor of Cas A. 
This is in addition to a recent estimate showing the progenitor to have been sub-solar \citep{2020ApJ...893...49S}.
However, \citet{2008ApJ...686..399S} constrained the maximum lifetime of a potential WR progenitor for Cas A to be $\lesssim 2000$ yrs, making the progenitor incompatible with EB SNe IIb.
Note that there is no companion to Cas A to deep limits \citep{2018MNRAS.473.1633K,2019A&A...623A..34K}. 

\section{Observationally Valuable Regimes for Constraining SN II\lowercase{b} Progenitors} \label{s:oppor}

While our models are able to account for current observational constraints for SNe IIb, there are regions favored by our models that are not represented in observations. Unobserved regimes either exclude the corresponding evolutionary scenarios or could be a result of observational bias.

{\it The only channel that can be excluded due to non-representation in observations is single and mildly interacting binaries.}
However, the remaining unobserved observational regimes can be attributed to observational bias and offer opportunities to target future observations for further constraining SN IIb progenitor channels.

The main channel that contributes to unobserved observational regimes is due to EB SNe IIb. 
They produce highly stripped and compact progenitors that are currently unobserved by all three observational probes considered in this work.
Specifically, our models favor the existence of SN IIb progenitors that are blue ($T_{\rm eff, preSN(,1)} \lesssim 10^5$K) with $\sim10^{4.5}-10^{5.5}~L_\sun$, radii $<50 R_\sun$ and low CSM densities.
Moreover, since these progenitors lie along the helium MS, their luminosities are correlated with their effective temperatures, which can aid in narrowing down follow-up strategies.
Detection of this group of SN IIb progenitors, especially in high metallicity environments, would favor the presence of weak stellar winds.
In fact, as we suggest in paper I, they are likely to exist even in high metallicity environments as they can account for observed SN IIb fractions.
The non-detection of this group of progenitors would indicate other channels toward SNe IIb \citep[e.g.][]{2006ApJ...640..891Y,2020arXiv200805076H}.
We note that, as discussed in Section \ref{s:models}, it is unclear whether explosions of compact EB SN IIb will result in a SN IIb or a SN Ib spectral classification.
The classification of IIb vs Ib can be observationally biased by the epoch when spectra are first obtained \citep{2011ApJ...739...41C,2013ApJ...767...71M}.
Interestingly, the photometric properties of EB SNe IIb progenitors are consistent with the identified progenitor of SN Ib iPTF13bvn. 
Moreover, the CSM properties of EB SNe IIb are consistent with those of SNe Ib \citep{2016ApJ...821...57D,2018ApJ...864...45M}.

Our models predict the existence of a dominant population of blue (MS) companions to SN IIb progenitors with $\sim10^{2.7}-10^{5.2}~L_\sun$ along with a smaller population of red and yellow (at high metallicity and mass transfer efficiency) companions with $\sim10^{4.4}-10^{5.4}~L_\sun$. 
The H-R space for companions can also help constrain the efficiency of mass transfer in binaries.
Specifically the existence of red evolved companions and low luminosity companions with $\sim10^{4.4}-10^{5.4}~L_\sun$ would favor inefficient mass transfer.
In fact, as we conclude in paper I, low mass transfer efficiencies ($\lesssim 0.1$) in addition to sub-solar stellar winds are needed, regardless of environment, to account for observed SN IIb fractions.

\section{Conclusions} \label{s:conclusions}

In this second paper of a two-part series aimed at conducting a comprehensive investigation of SNe IIb progenitors, 
we compare theoretical distributions of SN IIb progenitor properties to constraints from three independent observational probes.

Our models are successful in explaining the range in current observations.
We also identify observational regimes that either exclude some evolutionary scenarios or are potentially unobserved due to bias and can be targeted to further constrain progenitor channels (see Section \ref{s:oppor}).
The detection of SN IIb progenitors in these regimes would result in improved convergence of our understanding of SN progenitors from all observational lines of evidence: SN IIb fractions, progenitor structure constraints from LCs, direct progenitor detections in archival images, and CSM properties from X-ray/radio observations.
It would provide strong support for binaries as SN IIb progenitors experiencing weak stellar winds and inefficient mass transfer regardless of environment.

This paper demonstrates the importance of statistically comparing inferences from population scale modeling to observations.
However, there is currently a lack of statistical constraints for SN IIb populations from the observational side.
As mentioned in paper I, robust estimates of SN IIb fractions as function of metallicity will help improve insights from comparing to theoretical models.
There is also a need for distributions of constraints from LCs, ideally also as a function of metallicity.
The wealth of data from surveys like
All-Sky Automated Survey for Supernovae \citep{2017PASP..129j4502K}, Zwicky Transient Factory \citep{2014htu..conf...27B, 2017NatAs...1E..71B}, and, at the turn of the decade, LSST \citep{2002SPIE.4836...10T,2008arXiv0805.2366I} 
should help address both limitations.
It is harder to obtain large samples of direct progenitor detections because of the limited sensitivity of instruments (e.g. the detection horizon with HST is $\sim20$Mpc) and availability of pre-SN observations in relevant bands.
Moreover, while HST is expected to remain operational until at least 2025, there are currently no missions with the ability to undertake high-resolution UV imaging to aid in companion searches or for acquiring data that could lead to serendipitous hot progenitor (such as those favored in this work) identifications.
However, two of four 2020 Decadal Survey Mission Concept studies, LUVOIR \citep{2018arXiv180909668T} and HabEx \citep{2016SPIE.9904E..0LM}, can serve as successors to HST with significantly improved capabilities.
For example, LUVOIR can reach $\sim4$ times the depth of HST in $\sim10$\% the time \citep{2019arXiv191206219T}.
On the other hand, efforts toward conducting large sample studies of CSM properties are under way \citep[e.g.,][]{2016ApJ...821...57D,2018ApJ...864...45M}.
The availability of data discussed above along with theoretical models using accurate mass loss rates will lead to true convergence of our understanding of SN IIb progenitors from both theoretical and observational fronts.

\acknowledgments

We thank the referee for a close reading of the manuscript and helpful suggestions.
PM acknowledges support from the FWO junior postdoctoral fellowship No. 12ZY520N.
VK acknowledges support by the Canadian Institute for Advanced Research as a Fellow.
VK’s effort was supported through her CIFAR Senior Fellowship.
DM acknowledges NSF support from grants PHY-1914448 and AST-2037297.
NS thanks Ori Fox for inspiring thinking surrounding observational considerations for SN IIb progenitors.
This research was supported by computational resources provided for the Quest high performance computing facility at Northwestern University which is jointly supported by the National Science Foundation through grant NSF PHY-1126812, the Office of the Provost, the Office for Research, and Northwestern University Information Technology.
This research was supported in part through computational resources provided by Information Technology at Purdue, West Lafayette, Indiana.

\software{\mesa\ \citep[Release 9575][]{2011ApJS..192....3P,2013ApJS..208....4P,2015ApJS..220...15P,2018ApJS..234...34P}}

\newpage

\bibliography{references}

\begin{thebibliography}{}
\expandafter\ifx\csname natexlab\endcsname\relax\def\natexlab#1{#1}\fi
\providecommand{\url}[1]{\href{#1}{#1}}
\providecommand{\dodoi}[1]{doi:~\href{http://doi.org/#1}{\nolinkurl{#1}}}
\providecommand{\doeprint}[1]{\href{http://ascl.net/#1}{\nolinkurl{http://ascl.net/#1}}}
\providecommand{\doarXiv}[1]{\href{https://arxiv.org/abs/#1}{\nolinkurl{https://arxiv.org/abs/#1}}}

\bibitem[{{Aldering} {et~al.}(1994){Aldering}, {Humphreys}, \&
  {Richmond}}]{1994AJ....107..662A}
{Aldering}, G., {Humphreys}, R.~M., \& {Richmond}, M. 1994, \aj, 107, 662,
  \dodoi{10.1086/116886}

\bibitem[{{Arcavi} {et~al.}(2017){Arcavi}, {Hosseinzadeh}, {Brown}, {Smartt},
  {Valenti}, {Tartaglia}, {Piro}, {Sanchez}, {Nicholls}, {Monard}, {Howell},
  {McCully}, {Sand}, {Tonry}, {Denneau}, {Stalder}, {Heinze}, {Rest}, {Smith},
  \& {Bishop}}]{2017ApJ...837L...2A}
{Arcavi}, I., {Hosseinzadeh}, G., {Brown}, P.~J., {et~al.} 2017, \apjl, 837,
  L2, \dodoi{10.3847/2041-8213/aa5be1}

\bibitem[{{Arnett}(1982)}]{1982ApJ...253..785A}
{Arnett}, W.~D. 1982, \apj, 253, 785, \dodoi{10.1086/159681}

\bibitem[{{Arnett} \& {Meakin}(2011)}]{2011ApJ...741...33A}
{Arnett}, W.~D., \& {Meakin}, C. 2011, \apj, 741, 33,
  \dodoi{10.1088/0004-637X/741/1/33}

\bibitem[{{Bartunov} {et~al.}(1994){Bartunov}, {Blinnikov}, {Pavlyuk}, \&
  {Tsvetkov}}]{1994A&A...281L..53B}
{Bartunov}, O.~S., {Blinnikov}, S.~I., {Pavlyuk}, N.~N., \& {Tsvetkov}, D.~Y.
  1994, \aap, 281, L53

\bibitem[{{Bellm}(2014)}]{2014htu..conf...27B}
{Bellm}, E. 2014, in The Third Hot-wiring the Transient Universe Workshop, ed.
  P.~R. {Wozniak}, M.~J. {Graham}, A.~A. {Mahabal}, \& R.~{Seaman}, 27--33.
\newblock \doarXiv{1410.8185}

\bibitem[{{Bellm} \& {Kulkarni}(2017)}]{2017NatAs...1E..71B}
{Bellm}, E., \& {Kulkarni}, S. 2017, Nature Astronomy, 1, 0071,
  \dodoi{10.1038/s41550-017-0071}

\bibitem[{{Benvenuto} {et~al.}(2013){Benvenuto}, {Bersten}, \&
  {Nomoto}}]{2013ApJ...762...74B}
{Benvenuto}, O.~G., {Bersten}, M.~C., \& {Nomoto}, K. 2013, \apj, 762, 74,
  \dodoi{10.1088/0004-637X/762/2/74}

\bibitem[{{Bersten} {et~al.}(2012){Bersten}, {Benvenuto}, {Nomoto}, {Ergon},
  {Folatelli}, {Sollerman}, {Benetti}, {Botticella}, {Fraser}, {Kotak},
  {Maeda}, {Ochner}, \& {Tomasella}}]{2012ApJ...757...31B}
{Bersten}, M.~C., {Benvenuto}, O.~G., {Nomoto}, K., {et~al.} 2012, \apj, 757,
  31, \dodoi{10.1088/0004-637X/757/1/31}

\bibitem[{{Bersten} {et~al.}(2018){Bersten}, {Folatelli}, {Garc{\'{\i}}a}, {van
  Dyk}, {Benvenuto}, {Orellana}, {Buso}, {S{\'a}nchez}, {Tanaka}, {Maeda},
  {Filippenko}, {Zheng}, {Brink}, {Cenko}, {de Jaeger}, {Kumar}, {Moriya},
  {Nomoto}, {Perley}, {Shivvers}, \& {Smith}}]{2018Natur.554..497B}
{Bersten}, M.~C., {Folatelli}, G., {Garc{\'{\i}}a}, F., {et~al.} 2018, \nat,
  554, 497, \dodoi{10.1038/nature25151}

\bibitem[{{Bj{\"o}rklund} {et~al.}(2020){Bj{\"o}rklund}, {Sundqvist}, {Puls},
  \& {Najarro}}]{2020arXiv200806066B}
{Bj{\"o}rklund}, R., {Sundqvist}, J.~O., {Puls}, J., \& {Najarro}, F. 2020,
  arXiv e-prints, arXiv:2008.06066.
\newblock \doarXiv{2008.06066}

\bibitem[{{Brott} {et~al.}(2011){Brott}, {de Mink}, {Cantiello}, {Langer}, {de
  Koter}, {Evans}, {Hunter}, {Trundle}, \& {Vink}}]{2011A&A...530A.115B}
{Brott}, I., {de Mink}, S.~E., {Cantiello}, M., {et~al.} 2011, \aap, 530, A115,
  \dodoi{10.1051/0004-6361/201016113}

\bibitem[{{Bufano} {et~al.}(2014){Bufano}, {Pignata}, {Bersten}, {Mazzali},
  {Ryder}, {Margutti}, {Milisavljevic}, {Morelli}, {Benetti}, {Cappellaro},
  {Gonzalez-Gaitan}, {Romero-Ca{\~n}izales}, {Stritzinger}, {Walker},
  {Anderson}, {Contreras}, {de Jaeger}, {F{\"o}rster}, {Gutierrez}, {Hamuy},
  {Hsiao}, {Morrell}, {Olivares E.}, {Paillas}, {Parker}, {Pian}, {Pickering},
  {Sanders}, {Stockdale}, {Turatto}, {Valenti}, {Fesen}, {Maza}, {Nomoto},
  {Phillips}, \& {Soderberg}}]{2014MNRAS.439.1807B}
{Bufano}, F., {Pignata}, G., {Bersten}, M., {et~al.} 2014, \mnras, 439, 1807,
  \dodoi{10.1093/mnras/stu065}

\bibitem[{{Cao} {et~al.}(2013){Cao}, {Kasliwal}, {Arcavi}, {Horesh}, {Hancock},
  {Valenti}, {Cenko}, {Kulkarni}, {Gal-Yam}, {Gorbikov}, {Ofek}, {Sand},
  {Yaron}, {Graham}, {Silverman}, {Wheeler}, {Marion}, {Walker}, {Mazzali},
  {Howell}, {Li}, {Kong}, {Bloom}, {Nugent}, {Surace}, {Masci}, {Carpenter},
  {Degenaar}, \& {Gelino}}]{2013ApJ...775L...7C}
{Cao}, Y., {Kasliwal}, M.~M., {Arcavi}, I., {et~al.} 2013, \apjl, 775, L7,
  \dodoi{10.1088/2041-8205/775/1/L7}

\bibitem[{{Chornock} {et~al.}(2011){Chornock}, {Filippenko}, {Li}, {Marion},
  {Foley}, {Modjaz}, {Rafelski}, {Becker}, {de Vries}, {Garnavich},
  {Jorgenson}, {Lynch}, {Malec}, {Moran}, {Murphy}, {Rudy}, {Russell},
  {Silverman}, {Steele}, {Stockton}, {Wolfe}, \&
  {Woodward}}]{2011ApJ...739...41C}
{Chornock}, R., {Filippenko}, A.~V., {Li}, W., {et~al.} 2011, \apj, 739, 41,
  \dodoi{10.1088/0004-637X/739/1/41}

\bibitem[{{Claeys} {et~al.}(2011){Claeys}, {de Mink}, {Pols}, {Eldridge}, \&
  {Baes}}]{2011A&A...528A.131C}
{Claeys}, J.~S.~W., {de Mink}, S.~E., {Pols}, O.~R., {Eldridge}, J.~J., \&
  {Baes}, M. 2011, \aap, 528, A131, \dodoi{10.1051/0004-6361/201015410}

\bibitem[{{Crockett} {et~al.}(2008){Crockett}, {Eldridge}, {Smartt},
  {Pastorello}, {Gal-Yam}, {Fox}, {Leonard}, {Kasliwal}, {Mattila}, {Maund},
  {Stephens}, \& {Danziger}}]{2008MNRAS.391L...5C}
{Crockett}, R.~M., {Eldridge}, J.~J., {Smartt}, S.~J., {et~al.} 2008, \mnras,
  391, L5, \dodoi{10.1111/j.1745-3933.2008.00540.x}

\bibitem[{{de Jager} {et~al.}(1988){de Jager}, {Nieuwenhuijzen}, \& {van der
  Hucht}}]{1988A&AS...72..259D}
{de Jager}, C., {Nieuwenhuijzen}, H., \& {van der Hucht}, K.~A. 1988, \aaps,
  72, 259

\bibitem[{{Dessart} {et~al.}(2011){Dessart}, {Hillier}, {Livne}, {Yoon},
  {Woosley}, {Waldman}, \& {Langer}}]{2011MNRAS.414.2985D}
{Dessart}, L., {Hillier}, D.~J., {Livne}, E., {et~al.} 2011, \mnras, 414, 2985,
  \dodoi{10.1111/j.1365-2966.2011.18598.x}

\bibitem[{{Drout} {et~al.}(2016){Drout}, {Milisavljevic}, {Parrent},
  {Margutti}, {Kamble}, {Soderberg}, {Challis}, {Chornock}, {Fong}, {Frank},
  {Gehrels}, {Graham}, {Hsiao}, {Itagaki}, {Kasliwal}, {Kirshner}, {Macomb},
  {Marion}, {Norris}, \& {Phillips}}]{2016ApJ...821...57D}
{Drout}, M.~R., {Milisavljevic}, D., {Parrent}, J., {et~al.} 2016, \apj, 821,
  57, \dodoi{10.3847/0004-637X/821/1/57}

\bibitem[{{Ergon} {et~al.}(2015){Ergon}, {Jerkstrand}, {Sollerman},
  {Elias-Rosa}, {Fransson}, {Fraser}, {Pastorello}, {Kotak}, {Taubenberger},
  {Tomasella}, {Valenti}, {Benetti}, {Helou}, {Kasliwal}, {Maund}, {Smartt}, \&
  {Spyromilio}}]{Ergon2015}
{Ergon}, M., {Jerkstrand}, A., {Sollerman}, J., {et~al.} 2015, \aap, 580, A142,
  \dodoi{10.1051/0004-6361/201424592}

\bibitem[{{Folatelli} {et~al.}(2015){Folatelli}, {Bersten}, {Kuncarayakti},
  {Benvenuto}, {Maeda}, \& {Nomoto}}]{2015ApJ...811..147F}
{Folatelli}, G., {Bersten}, M.~C., {Kuncarayakti}, H., {et~al.} 2015, \apj,
  811, 147, \dodoi{10.1088/0004-637X/811/2/147}

\bibitem[{{Folatelli} {et~al.}(2014){Folatelli}, {Bersten}, {Benvenuto}, {Van
  Dyk}, {Kuncarayakti}, {Maeda}, {Nozawa}, {Nomoto}, {Hamuy}, \&
  {Quimby}}]{2014ApJ...793L..22F}
{Folatelli}, G., {Bersten}, M.~C., {Benvenuto}, O.~G., {et~al.} 2014, \apjl,
  793, L22, \dodoi{10.1088/2041-8205/793/2/L22}

\bibitem[{{Folatelli} {et~al.}(2016){Folatelli}, {Van Dyk}, {Kuncarayakti},
  {Maeda}, {Bersten}, {Nomoto}, {Pignata}, {Hamuy}, {Quimby}, {Zheng},
  {Filippenko}, {Clubb}, {Smith}, {Elias-Rosa}, {Foley}, \&
  {Miller}}]{2016ApJ...825L..22F}
{Folatelli}, G., {Van Dyk}, S.~D., {Kuncarayakti}, H., {et~al.} 2016, \apjl,
  825, L22, \dodoi{10.3847/2041-8205/825/2/L22}

\bibitem[{{Fox} {et~al.}(2014){Fox}, {Azalee Bostroem}, {Van Dyk},
  {Filippenko}, {Fransson}, {Matheson}, {Cenko}, {Chandra}, {Dwarkadas}, {Li},
  {Parker}, \& {Smith}}]{2014ApJ...790...17F}
{Fox}, O.~D., {Azalee Bostroem}, K., {Van Dyk}, S.~D., {et~al.} 2014, \apj,
  790, 17, \dodoi{10.1088/0004-637X/790/1/17}

\bibitem[{{Fransson} {et~al.}(1996){Fransson}, {Lundqvist}, \&
  {Chevalier}}]{1996ApJ...461..993F}
{Fransson}, C., {Lundqvist}, P., \& {Chevalier}, R.~A. 1996, \apj, 461, 993,
  \dodoi{10.1086/177119}

\bibitem[{{Fremling} {et~al.}(2016){Fremling}, {Sollerman}, {Taddia}, {Ergon},
  {Fraser}, {Karamehmetoglu}, {Valenti}, {Jerkstrand}, {Arcavi}, {Bufano},
  {Elias Rosa}, {Filippenko}, {Fox}, {Gal-Yam}, {Howell}, {Kotak}, {Mazzali},
  {Milisavljevic}, {Nugent}, {Nyholm}, {Pian}, \& {Smartt}}]{Fremling2016}
{Fremling}, C., {Sollerman}, J., {Taddia}, F., {et~al.} 2016, \aap, 593, A68,
  \dodoi{10.1051/0004-6361/201628275}

\bibitem[{{Fremling} {et~al.}(2019){Fremling}, {Ko}, {Dugas}, {Ergon},
  {Sollerman}, {Bagdasaryan}, {Barbarino}, {Belicki}, {Bellm}, {Blagorodnova},
  {De}, {Dekany}, {Frederick}, {Gal-Yam}, {Goldstein}, {Golkhou}, {Graham},
  {Kasliwal}, {Kowalski}, {Kulkarni}, {Kupfer}, {Laher}, {Masci}, {Miller},
  {Neill}, {Perley}, {Rebbapragada}, {Riddle}, {Rusholme}, {Schulze}, {Smith},
  {Tartaglia}, {Yan}, \& {Yao}}]{2019ApJ...878L...5F}
{Fremling}, C., {Ko}, H., {Dugas}, A., {et~al.} 2019, \apjl, 878, L5,
  \dodoi{10.3847/2041-8213/ab218f}

\bibitem[{{Gal-Yam} {et~al.}(2014){Gal-Yam}, {Arcavi}, {Ofek}, {Ben-Ami},
  {Cenko}, {Kasliwal}, {Cao}, {Yaron}, {Tal}, {Silverman}, {Horesh}, {De Cia},
  {Taddia}, {Sollerman}, {Perley}, {Vreeswijk}, {Kulkarni}, {Nugent},
  {Filippenko}, \& {Wheeler}}]{2014Natur.509..471G}
{Gal-Yam}, A., {Arcavi}, I., {Ofek}, E.~O., {et~al.} 2014, \nat, 509, 471,
  \dodoi{10.1038/nature13304}

\bibitem[{{Gangopadhyay} {et~al.}(2018){Gangopadhyay}, {Misra}, {Pastorello},
  {Sahu}, {Tomasella}, {Tartaglia}, {Singh}, {Dastidar}, {Srivastav}, {Ochner},
  {Brown}, {Anupama}, {Benetti}, {Cappellaro}, {Kumar}, {Kumar}, \&
  {Pandey}}]{2018MNRAS.476.3611G}
{Gangopadhyay}, A., {Misra}, K., {Pastorello}, A., {et~al.} 2018, \mnras, 476,
  3611, \dodoi{10.1093/mnras/sty478}

\bibitem[{{Georgy} {et~al.}(2012){Georgy}, {Ekstr{\"o}m}, {Meynet}, {Massey},
  {Levesque}, {Hirschi}, {Eggenberger}, \& {Maeder}}]{2012A&A...542A..29G}
{Georgy}, C., {Ekstr{\"o}m}, S., {Meynet}, G., {et~al.} 2012, \aap, 542, A29,
  \dodoi{10.1051/0004-6361/201118340}

\bibitem[{{Goldman} {et~al.}(2017){Goldman}, {van Loon}, {Zijlstra}, {Green},
  {Wood}, {Nanni}, {Imai}, {Whitelock}, {Matsuura}, {Groenewegen}, \&
  {G{\'o}mez}}]{2017MNRAS.465..403G}
{Goldman}, S.~R., {van Loon}, J.~T., {Zijlstra}, A.~A., {et~al.} 2017, \mnras,
  465, 403, \dodoi{10.1093/mnras/stw2708}

\bibitem[{{G{\"o}tberg} {et~al.}(2018){G{\"o}tberg}, {de Mink}, {Groh},
  {Kupfer}, {Crowther}, {Zapartas}, \& {Renzo}}]{2018A&A...615A..78G}
{G{\"o}tberg}, Y., {de Mink}, S.~E., {Groh}, J.~H., {et~al.} 2018, \aap, 615,
  A78, \dodoi{10.1051/0004-6361/201732274}

\bibitem[{{Groh}(2014)}]{2014A&A...572L..11G}
{Groh}, J.~H. 2014, \aap, 572, L11, \dodoi{10.1051/0004-6361/201424852}

\bibitem[{{Groh} {et~al.}(2013{\natexlab{a}}){Groh}, {Meynet}, \&
  {Ekstr{\"o}m}}]{2013A&A...550L...7G}
{Groh}, J.~H., {Meynet}, G., \& {Ekstr{\"o}m}, S. 2013{\natexlab{a}}, \aap,
  550, L7, \dodoi{10.1051/0004-6361/201220741}

\bibitem[{{Groh} {et~al.}(2013{\natexlab{b}}){Groh}, {Meynet}, {Georgy}, \&
  {Ekstr{\"o}m}}]{2013A&A...558A.131G}
{Groh}, J.~H., {Meynet}, G., {Georgy}, C., \& {Ekstr{\"o}m}, S.
  2013{\natexlab{b}}, \aap, 558, A131, \dodoi{10.1051/0004-6361/201321906}

\bibitem[{{Hirai} {et~al.}(2020){Hirai}, {Sato}, {Podsiadlowski},
  {Vigna-Gomez}, \& {Mandel}}]{2020arXiv200805076H}
{Hirai}, R., {Sato}, T., {Podsiadlowski}, P., {Vigna-Gomez}, A., \& {Mandel},
  I. 2020, arXiv e-prints, arXiv:2008.05076.
\newblock \doarXiv{2008.05076}

\bibitem[{{Houck} \& {Fransson}(1996)}]{1996ApJ...456..811H}
{Houck}, J.~C., \& {Fransson}, C. 1996, \apj, 456, 811, \dodoi{10.1086/176699}

\bibitem[{{Iglesias} \& {Rogers}(1996)}]{1996ApJ...464..943I}
{Iglesias}, C.~A., \& {Rogers}, F.~J. 1996, \apj, 464, 943,
  \dodoi{10.1086/177381}

\bibitem[{{Ivezic} {et~al.}(2008){Ivezic}, {Tyson}, {Abel}, {Acosta},
  {Allsman}, {AlSayyad}, {Anderson}, {Andrew}, {Angel}, {Angeli}, {Ansari},
  {Antilogus}, {Arndt}, {Astier}, {Aubourg}, {Axelrod}, {Bard}, {Barr},
  {Barrau}, {Bartlett}, {Bauman}, {Beaumont}, {Becker}, {Becla}, {Beldica},
  {Bellavia}, {Blanc}, {Blandford}, {Bloom}, {Bogart}, {Borne}, {Bosch},
  {Boutigny}, {Brandt}, {Brown}, {Bullock}, {Burchat}, {Burke}, {Cagnoli},
  {Calabrese}, {Chandrasekharan}, {Chesley}, {Cheu}, {Chiang}, {Claver},
  {Connolly}, {Cook}, {Cooray}, {Covey}, {Cribbs}, {Cui}, {Cutri}, {Daubard},
  {Daues}, {Delgado}, {Digel}, {Doherty}, {Dubois}, {Dubois-Felsmann},
  {Durech}, {Eracleous}, {Ferguson}, {Frank}, {Freemon}, {Gangler}, {Gawiser},
  {Geary}, {Gee}, {Geha}, {Gibson}, {Gilmore}, {Glanzman}, {Goodenow},
  {Gressler}, {Gris}, {Guyonnet}, {Hascall}, {Haupt}, {Hernandez}, {Hogan},
  {Huang}, {Huffer}, {Innes}, {Jacoby}, {Jain}, {Jee}, {Jernigan},
  {Jevremovic}, {Johns}, {Jones}, {Juramy-Gilles}, {Juric}, {Kahn}, {Kalirai},
  {Kallivayalil}, {Kalmbach}, {Kantor}, {Kasliwal}, {Kessler}, {Kirkby},
  {Knox}, {Kotov}, {Krabbendam}, {Krughoff}, {Kubanek}, {Kuczewski},
  {Kulkarni}, {Lambert}, {Le Guillou}, {Levine}, {Liang}, {Lim}, {Lintott},
  {Lupton}, {Mahabal}, {Marshall}, {Marshall}, {May}, {McKercher}, {Migliore},
  {Miller}, {Mills}, {Monet}, {Moniez}, {Neill}, {Nief}, {Nomerotski},
  {Nordby}, {O'Connor}, {Oliver}, {Olivier}, {Olsen}, {Ortiz}, {Owen}, {Pain},
  {Peterson}, {Petry}, {Pierfederici}, {Pietrowicz}, {Pike}, {Pinto}, {Plante},
  {Plate}, {Price}, {Prouza}, {Radeka}, {Rajagopal}, {Rasmussen}, {Regnault},
  {Ridgway}, {Ritz}, {Rosing}, {Roucelle}, {Rumore}, {Russo}, {Saha},
  {Sassolas}, {Schalk}, {Schindler}, {Schneider}, {Schumacher}, {Sebag},
  {Sembroski}, {Seppala}, {Shipsey}, {Silvestri}, {Smith}, {Smith}, {Strauss},
  {Stubbs}, {Sweeney}, {Szalay}, {Takacs}, {Thaler}, {Van Berg}, {Vanden Berk},
  {Vetter}, {Virieux}, {Xin}, {Walkowicz}, {Walter}, {Wang}, {Warner},
  {Willman}, {Wittman}, {Wolff}, {Wood-Vasey}, {Yoachim}, {Zhan}, \& {for the
  LSST Collaboration}}]{2008arXiv0805.2366I}
{Ivezic}, Z., {Tyson}, J.~A., {Abel}, B., {et~al.} 2008, ArXiv e-prints.
\newblock \doarXiv{0805.2366}

\bibitem[{{Kamble} {et~al.}(2016){Kamble}, {Margutti}, {Soderberg},
  {Chakraborti}, {Fransson}, {Chevalier}, {Powell}, {Milisavljevic}, {Parrent},
  \& {Bietenholz}}]{2016ApJ...818..111K}
{Kamble}, A., {Margutti}, R., {Soderberg}, A.~M., {et~al.} 2016, \apj, 818,
  111, \dodoi{10.3847/0004-637X/818/2/111}

\bibitem[{{Kerzendorf} {et~al.}(2019){Kerzendorf}, {Do}, {de Mink},
  {G{\"o}tberg}, {Milisavljevic}, {Zapartas}, {Renzo}, {Justham},
  {Podsiadlowski}, \& {Fesen}}]{2019A&A...623A..34K}
{Kerzendorf}, W.~E., {Do}, T., {de Mink}, S.~E., {et~al.} 2019, \aap, 623, A34,
  \dodoi{10.1051/0004-6361/201732206}

\bibitem[{{Kilpatrick} {et~al.}(2017){Kilpatrick}, {Foley}, {Abramson}, {Pan},
  {Lu}, {Williams}, {Treu}, {Siebert}, {Fassnacht}, \&
  {Max}}]{2017MNRAS.465.4650K}
{Kilpatrick}, C.~D., {Foley}, R.~J., {Abramson}, L.~E., {et~al.} 2017, \mnras,
  465, 4650, \dodoi{10.1093/mnras/stw3082}

\bibitem[{{Kilpatrick} {et~al.}(2018){Kilpatrick}, {Takaro}, {Foley},
  {Leibler}, {Pan}, {Campbell}, {Jacobson-Galan}, {Lewis}, {Lyke}, {Max},
  {Medallon}, \& {Rest}}]{2018MNRAS.480.2072K}
{Kilpatrick}, C.~D., {Takaro}, T., {Foley}, R.~J., {et~al.} 2018, \mnras, 480,
  2072, \dodoi{10.1093/mnras/sty2022}

\bibitem[{{Kippenhahn} {et~al.}(1980){Kippenhahn}, {Ruschenplatt}, \&
  {Thomas}}]{1980A&A....91..175K}
{Kippenhahn}, R., {Ruschenplatt}, G., \& {Thomas}, H.-C. 1980, \aap, 91, 175

\bibitem[{{Kobulnicky} {et~al.}(2014){Kobulnicky}, {Kiminki}, {Lundquist},
  {Burke}, {Chapman}, {Keller}, {Lester}, {Rolen}, {Topel}, {Bhattacharjee},
  {Smullen}, {Vargas {\'A}lvarez}, {Runnoe}, {Dale}, \&
  {Brotherton}}]{2014ApJS..213...34K}
{Kobulnicky}, H.~A., {Kiminki}, D.~C., {Lundquist}, M.~J., {et~al.} 2014,
  \apjs, 213, 34, \dodoi{10.1088/0067-0049/213/2/34}

\bibitem[{{Kochanek}(2018)}]{2018MNRAS.473.1633K}
{Kochanek}, C.~S. 2018, \mnras, 473, 1633, \dodoi{10.1093/mnras/stx2423}

\bibitem[{{Kochanek} {et~al.}(2017){Kochanek}, {Shappee}, {Stanek}, {Holoien},
  {Thompson}, {Prieto}, {Dong}, {Shields}, {Will}, {Britt}, {Perzanowski}, \&
  {Pojma{\'n}ski}}]{2017PASP..129j4502K}
{Kochanek}, C.~S., {Shappee}, B.~J., {Stanek}, K.~Z., {et~al.} 2017, \pasp,
  129, 104502, \dodoi{10.1088/1538-3873/aa80d9}

\bibitem[{{K{\"o}hler} {et~al.}(2015){K{\"o}hler}, {Langer}, {de Koter}, {de
  Mink}, {Crowther}, {Evans}, {Gr{\"a}fener}, {Sana}, {Sanyal}, {Schneider}, \&
  {Vink}}]{2015A&A...573A..71K}
{K{\"o}hler}, K., {Langer}, N., {de Koter}, A., {et~al.} 2015, \aap, 573, A71,
  \dodoi{10.1051/0004-6361/201424356}

\bibitem[{{Kolb} \& {Ritter}(1990)}]{1990A&A...236..385K}
{Kolb}, U., \& {Ritter}, H. 1990, \aap, 236, 385

\bibitem[{{Krauss} {et~al.}(2012){Krauss}, {Soderberg}, {Chomiuk}, {Zauderer},
  {Brunthaler}, {Bietenholz}, {Chevalier}, {Fransson}, \&
  {Rupen}}]{2012ApJ...750L..40K}
{Krauss}, M.~I., {Soderberg}, A.~M., {Chomiuk}, L., {et~al.} 2012, \apj, 750,
  L40, \dodoi{10.1088/2041-8205/750/2/L40}

\bibitem[{{Kroupa}(2001)}]{2001MNRAS.322..231K}
{Kroupa}, P. 2001, \mnras, 322, 231, \dodoi{10.1046/j.1365-8711.2001.04022.x}

\bibitem[{{Langer} {et~al.}(1983){Langer}, {Fricke}, \&
  {Sugimoto}}]{1983A&A...126..207L}
{Langer}, N., {Fricke}, K.~J., \& {Sugimoto}, D. 1983, \aap, 126, 207

\bibitem[{{Lohev} {et~al.}(2019){Lohev}, {Sabach}, {Gilkis}, \&
  {Soker}}]{2019MNRAS.490....9L}
{Lohev}, N., {Sabach}, E., {Gilkis}, A., \& {Soker}, N. 2019, \mnras, 490, 9,
  \dodoi{10.1093/mnras/stz2593}

\bibitem[{{Lyman} {et~al.}(2016){Lyman}, {Bersier}, {James}, {Mazzali},
  {Eldridge}, {Fraser}, \& {Pian}}]{2016MNRAS.457..328L}
{Lyman}, J.~D., {Bersier}, D., {James}, P.~A., {et~al.} 2016, \mnras, 457, 328,
  \dodoi{10.1093/mnras/stv2983}

\bibitem[{{Maeda} {et~al.}(2015){Maeda}, {Hattori}, {Milisavljevic},
  {Folatelli}, {Drout}, {Kuncarayakti}, {Margutti}, {Kamble}, {Soderberg},
  {Tanaka}, {Kawabata}, {Kawabata}, {Yamanaka}, {Nomoto}, {Kim}, {Simon},
  {Phillips}, {Parrent}, {Nakaoka}, {Moriya}, {Suzuki}, {Takaki}, {Ishigaki},
  {Sakon}, {Tajitsu}, \& {Iye}}]{2015ApJ...807...35M}
{Maeda}, K., {Hattori}, T., {Milisavljevic}, D., {et~al.} 2015, \apj, 807, 35,
  \dodoi{10.1088/0004-637X/807/1/35}

\bibitem[{{Margutti} {et~al.}(2017){Margutti}, {Kamble}, {Milisavljevic},
  {Zapartas}, {de Mink}, {Drout}, {Chornock}, {Risaliti}, {Zauderer},
  {Bietenholz}, {Cantiello}, {Chakraborti}, {Chomiuk}, {Fong}, {Grefenstette},
  {Guidorzi}, {Kirshner}, {Parrent}, {Patnaude}, {Soderberg}, {Gehrels}, \&
  {Harrison}}]{2017ApJ...835..140M}
{Margutti}, R., {Kamble}, A., {Milisavljevic}, D., {et~al.} 2017, \apj, 835,
  140, \dodoi{10.3847/1538-4357/835/2/140}

\bibitem[{{Margutti} {et~al.}(2018){Margutti}, {Chornock}, {Metzger},
  {Coppejans}, {Guidorzi}, {Migliori}, {Milisavljevic}, {Berger}, {Nicholl},
  {Zauderer}, {Lunnan}, {Kamble}, {Drout}, \& {Modjaz}}]{2018ApJ...864...45M}
{Margutti}, R., {Chornock}, R., {Metzger}, B.~D., {et~al.} 2018, \apj, 864, 45,
  \dodoi{10.3847/1538-4357/aad2df}

\bibitem[{{Maund} {et~al.}(2004){Maund}, {Smartt}, {Kudritzki},
  {Podsiadlowski}, \& {Gilmore}}]{2004Natur.427..129M}
{Maund}, J.~R., {Smartt}, S.~J., {Kudritzki}, R.~P., {Podsiadlowski}, P., \&
  {Gilmore}, G.~F. 2004, \nat, 427, 129

\bibitem[{{Maund} {et~al.}(2011){Maund}, {Fraser}, {Ergon}, {Pastorello},
  {Smartt}, {Sollerman}, {Benetti}, {Botticella}, {Bufano}, {Danziger},
  {Kotak}, {Magill}, {Stephens}, \& {Valenti}}]{2011ApJ...739L..37M}
{Maund}, J.~R., {Fraser}, M., {Ergon}, M., {et~al.} 2011, \apjl, 739, L37,
  \dodoi{10.1088/2041-8205/739/2/L37}

\bibitem[{{Mazzali} {et~al.}(2008){Mazzali}, {Valenti}, {Della Valle},
  {Chincarini}, {Sauer}, {Benetti}, {Pian}, {Piran}, {D'Elia}, {Elias-Rosa},
  {Margutti}, {Pasotti}, {Antonelli}, {Bufano}, {Campana}, {Cappellaro},
  {Covino}, {D'Avanzo}, {Fiore}, {Fugazza}, {Gilmozzi}, {Hunter}, {Maguire},
  {Maiorano}, {Marziani}, {Masetti}, {Mirabel}, {Navasardyan}, {Nomoto},
  {Palazzi}, {Pastorello}, {Panagia}, {Pellizza}, {Sari}, {Smartt},
  {Tagliaferri}, {Tanaka}, {Taubenberger}, {Tominaga}, {Trundle}, \&
  {Turatto}}]{2008Sci...321.1185M}
{Mazzali}, P.~A., {Valenti}, S., {Della Valle}, M., {et~al.} 2008, Science,
  321, 1185, \dodoi{10.1126/science.1158088}

\bibitem[{{Mennesson} {et~al.}(2016){Mennesson}, {Gaudi}, {Seager}, {Cahoy},
  {Domagal-Goldman}, {Feinberg}, {Guyon}, {Kasdin}, {Marois}, {Mawet},
  {Tamura}, {Mouillet}, {Prusti}, {Quirrenbach}, {Robinson}, {Rogers},
  {Scowen}, {Somerville}, {Stapelfeldt}, {Stern}, {Still}, {Turnbull}, {Booth},
  {Kiessling}, {Kuan}, \& {Warfield}}]{2016SPIE.9904E..0LM}
{Mennesson}, B., {Gaudi}, S., {Seager}, S., {et~al.} 2016, in Society of
  Photo-Optical Instrumentation Engineers (SPIE) Conference Series, Vol. 9904,
  \procspie, 99040L, \dodoi{10.1117/12.2240457}

\bibitem[{{Meynet} {et~al.}(2015){Meynet}, {Chomienne}, {Ekstr{\"o}m},
  {Georgy}, {Granada}, {Groh}, {Maeder}, {Eggenberger}, {Levesque}, \&
  {Massey}}]{2015A&A...575A..60M}
{Meynet}, G., {Chomienne}, V., {Ekstr{\"o}m}, S., {et~al.} 2015, \aap, 575,
  A60, \dodoi{10.1051/0004-6361/201424671}

\bibitem[{{Milisavljevic} {et~al.}(2013){Milisavljevic}, {Margutti},
  {Soderberg}, {Pignata}, {Chomiuk}, {Fesen}, {Bufano}, {Sanders}, {Parrent},
  {Parker}, {Mazzali}, {Pian}, {Pickering}, {Buckley}, {Crawford}, {Gulbis},
  {Hettlage}, {Hooper}, {Nordsieck}, {O'Donoghue}, {Husser}, {Potter},
  {Kniazev}, {Kotze}, {Romero-Colmenero}, {Vaisanen}, {Wolf}, {Bietenholz},
  {Bartel}, {Fransson}, {Walker}, {Brunthaler}, {Chakraborti}, {Levesque},
  {MacFadyen}, {Drescher}, {Bock}, {Marples}, {Anderson}, {Benetti},
  {Reichart}, \& {Ivarsen}}]{2013ApJ...767...71M}
{Milisavljevic}, D., {Margutti}, R., {Soderberg}, A.~M., {et~al.} 2013, \apj,
  767, 71, \dodoi{10.1088/0004-637X/767/1/71}

\bibitem[{{Milisavljevic} {et~al.}(2015){Milisavljevic}, {Margutti}, {Kamble},
  {Patnaude}, {Raymond}, {Eldridge}, {Fong}, {Bietenholz}, {Challis},
  {Chornock}, {Drout}, {Fransson}, {Fesen}, {Grindlay}, {Kirshner}, {Lunnan},
  {Mackey}, {Miller}, {Parrent}, {Sand ers}, {Soderberg}, \&
  {Zauderer}}]{2015ApJ...815..120M}
{Milisavljevic}, D., {Margutti}, R., {Kamble}, A., {et~al.} 2015, \apj, 815,
  120, \dodoi{10.1088/0004-637X/815/2/120}

\bibitem[{{Modjaz} {et~al.}(2009){Modjaz}, {Li}, {Butler}, {Chornock},
  {Perley}, {Blondin}, {Bloom}, {Filippenko}, {Kirshner}, {Kocevski},
  {Poznanski}, {Hicken}, {Foley}, {Stringfellow}, {Berlind}, {Barrado y
  Navascues}, {Blake}, {Bouy}, {Brown}, {Challis}, {Chen}, {de Vries},
  {Dufour}, {Falco}, {Friedman}, {Ganeshalingam}, {Garnavich}, {Holden},
  {Illingworth}, {Lee}, {Liebert}, {Marion}, {Olivier}, {Prochaska},
  {Silverman}, {Smith}, {Starr}, {Steele}, {Stockton}, {Williams}, \&
  {Wood-Vasey}}]{2009ApJ...702..226M}
{Modjaz}, M., {Li}, W., {Butler}, N., {et~al.} 2009, \apj, 702, 226,
  \dodoi{10.1088/0004-637X/702/1/226}

\bibitem[{{Moe} \& {Di Stefano}(2017)}]{2017ApJS..230...15M}
{Moe}, M., \& {Di Stefano}, R. 2017, \apjs, 230, 15,
  \dodoi{10.3847/1538-4365/aa6fb6}

\bibitem[{{Morales-Garoffolo} {et~al.}(2014){Morales-Garoffolo}, {Elias-Rosa},
  {Benetti}, {Taubenberger}, {Cappellaro}, {Pastorello}, {Klauser}, {Valenti},
  {Howerton}, {Ochner}, {Schramm}, {Siviero}, {Tartaglia}, \&
  {Tomasella}}]{2014MNRAS.445.1647M}
{Morales-Garoffolo}, A., {Elias-Rosa}, N., {Benetti}, S., {et~al.} 2014,
  \mnras, 445, 1647, \dodoi{10.1093/mnras/stu1837}

\bibitem[{{Morales-Garoffolo} {et~al.}(2015){Morales-Garoffolo}, {Elias-Rosa},
  {Bersten}, {Jerkstrand}, {Taubenberger}, {Benetti}, {Cappellaro}, {Kotak},
  {Pastorello}, {Bufano}, {Dom{\'{\i}}nguez}, {Ergon}, {Fraser}, {Gao},
  {Garc{\'{\i}}a}, {Howell}, {Isern}, {Smartt}, {Tomasella}, \&
  {Valenti}}]{2015MNRAS.454...95M}
{Morales-Garoffolo}, A., {Elias-Rosa}, N., {Bersten}, M., {et~al.} 2015,
  \mnras, 454, 95, \dodoi{10.1093/mnras/stv1972}

\bibitem[{{Moriya} {et~al.}(2016){Moriya}, {Pruzhinskaya}, {Ergon}, \&
  {Blinnikov}}]{2016MNRAS.455..423M}
{Moriya}, T.~J., {Pruzhinskaya}, M.~V., {Ergon}, M., \& {Blinnikov}, S.~I.
  2016, \mnras, 455, 423, \dodoi{10.1093/mnras/stv2336}

\bibitem[{{Nagy} \& {Vink{\'o}}(2016)}]{2016A&A...589A..53N}
{Nagy}, A.~P., \& {Vink{\'o}}, J. 2016, \aap, 589, A53,
  \dodoi{10.1051/0004-6361/201527931}

\bibitem[{{Nakar} \& {Piro}(2014)}]{2014ApJ...788..193N}
{Nakar}, E., \& {Piro}, A.~L. 2014, \apj, 788, 193,
  \dodoi{10.1088/0004-637X/788/2/193}

\bibitem[{{Nomoto} {et~al.}(1993){Nomoto}, {Suzuki}, {Shigeyama}, {Kumagai},
  {Yamaoka}, \& {Saio}}]{1993Natur.364..507N}
{Nomoto}, K., {Suzuki}, T., {Shigeyama}, T., {et~al.} 1993, \nat, 364, 507,
  \dodoi{10.1038/364507a0}

\bibitem[{{Nugis} \& {Lamers}(2000)}]{2000A&A...360..227N}
{Nugis}, T., \& {Lamers}, H.~J.~G.~L.~M. 2000, \aap, 360, 227

\bibitem[{{Oates} {et~al.}(2012){Oates}, {Bayless}, {Stritzinger}, {Prichard},
  {Prieto}, {Immler}, {Brown}, {Breeveld}, {De Pasquale}, {Kuin}, {Hamuy},
  {Holland}, {Taddia}, \& {Roming}}]{2012MNRAS.424.1297O}
{Oates}, S.~R., {Bayless}, A.~J., {Stritzinger}, M.~D., {et~al.} 2012, \mnras,
  424, 1297, \dodoi{10.1111/j.1365-2966.2012.21311.x}

\bibitem[{{Ouchi} \& {Maeda}(2017)}]{2017ApJ...840...90O}
{Ouchi}, R., \& {Maeda}, K. 2017, \apj, 840, 90,
  \dodoi{10.3847/1538-4357/aa6ea9}

\bibitem[{{Packet}(1981)}]{1981A&A...102...17P}
{Packet}, W. 1981, \aap, 102, 17

\bibitem[{{Paxton} {et~al.}(2011){Paxton}, {Bildsten}, {Dotter}, {Herwig},
  {Lesaffre}, \& {Timmes}}]{2011ApJS..192....3P}
{Paxton}, B., {Bildsten}, L., {Dotter}, A., {et~al.} 2011, \apjs, 192, 3,
  \dodoi{10.1088/0067-0049/192/1/3}

\bibitem[{{Paxton} {et~al.}(2013){Paxton}, {Cantiello}, {Arras}, {Bildsten},
  {Brown}, {Dotter}, {Mankovich}, {Montgomery}, {Stello}, {Timmes}, \&
  {Townsend}}]{2013ApJS..208....4P}
{Paxton}, B., {Cantiello}, M., {Arras}, P., {et~al.} 2013, \apjs, 208, 4,
  \dodoi{10.1088/0067-0049/208/1/4}

\bibitem[{{Paxton} {et~al.}(2015){Paxton}, {Marchant}, {Schwab}, {Bauer},
  {Bildsten}, {Cantiello}, {Dessart}, {Farmer}, {Hu}, {Langer}, {Townsend},
  {Townsley}, \& {Timmes}}]{2015ApJS..220...15P}
{Paxton}, B., {Marchant}, P., {Schwab}, J., {et~al.} 2015, \apjs, 220, 15,
  \dodoi{10.1088/0067-0049/220/1/15}

\bibitem[{{Paxton} {et~al.}(2018){Paxton}, {Schwab}, {Bauer}, {Bildsten},
  {Blinnikov}, {Duffell}, {Farmer}, {Goldberg}, {Marchant}, {Sorokina},
  {Thoul}, {Townsend}, \& {Timmes}}]{2018ApJS..234...34P}
{Paxton}, B., {Schwab}, J., {Bauer}, E.~B., {et~al.} 2018, \apjs, 234, 34,
  \dodoi{10.3847/1538-4365/aaa5a8}

\bibitem[{{Peimbert} {et~al.}(2007){Peimbert}, {Luridiana}, \&
  {Peimbert}}]{2007ApJ...666..636P}
{Peimbert}, M., {Luridiana}, V., \& {Peimbert}, A. 2007, \apj, 666, 636,
  \dodoi{10.1086/520571}

\bibitem[{{P{\'e}rez-Torres} {et~al.}(2005){P{\'e}rez-Torres}, {Alberdi},
  {Marcaide}, {Guerrero}, {Lundqvist}, {Shapiro}, {Ros}, {Lara}, {Guirado},
  {Weiler}, \& {Stockdale}}]{2005MNRAS.360.1055P}
{P{\'e}rez-Torres}, M.~A., {Alberdi}, A., {Marcaide}, J.~M., {et~al.} 2005,
  \mnras, 360, 1055, \dodoi{10.1111/j.1365-2966.2005.09102.x}

\bibitem[{{Petrovic} {et~al.}(2005){Petrovic}, {Langer}, \& {van der
  Hucht}}]{2005A&A...435.1013P}
{Petrovic}, J., {Langer}, N., \& {van der Hucht}, K.~A. 2005, \aap, 435, 1013,
  \dodoi{10.1051/0004-6361:20042368}

\bibitem[{{Piro}(2015)}]{2015ApJ...808L..51P}
{Piro}, A.~L. 2015, \apj, 808, L51, \dodoi{10.1088/2041-8205/808/2/L51}

\bibitem[{{Podsiadlowski} {et~al.}(1993){Podsiadlowski}, {Hsu}, {Joss}, \&
  {Ross}}]{1993Natur.364..509P}
{Podsiadlowski}, P., {Hsu}, J.~J.~L., {Joss}, P.~C., \& {Ross}, R.~R. 1993,
  \nat, 364, 509, \dodoi{10.1038/364509a0}

\bibitem[{{Podsiadlowski} {et~al.}(1992){Podsiadlowski}, {Joss}, \&
  {Hsu}}]{1992ApJ...391..246P}
{Podsiadlowski}, P., {Joss}, P.~C., \& {Hsu}, J.~J.~L. 1992, \apj, 391, 246,
  \dodoi{10.1086/171341}

\bibitem[{{Popham} \& {Narayan}(1991)}]{1991ApJ...370..604P}
{Popham}, R., \& {Narayan}, R. 1991, \apj, 370, 604, \dodoi{10.1086/169847}

\bibitem[{{Prentice} {et~al.}(2019){Prentice}, {Ashall}, {James}, {Short},
  {Mazzali}, {Bersier}, {Crowther}, {Barbarino}, {Chen}, {Copperwheat},
  {Darnley}, {Denneau}, {Elias-Rosa}, {Fraser}, {Galbany}, {Gal-Yam},
  {Harmanen}, {Howell}, {Hosseinzadeh}, {Inserra}, {Kankare}, {Karamehmetoglu},
  {Lamb}, {Limongi}, {Maguire}, {McCully}, {Olivares E}, {Piascik}, {Pignata},
  {Reichart}, {Rest}, {Reynolds}, {Rodr{\'\i}guez}, {Saario}, {Schulze},
  {Smartt}, {Smith}, {Sollerman}, {Stalder}, {Sullivan}, {Taddia}, {Valenti},
  {Vergani}, {Williams}, \& {Young}}]{2019MNRAS.485.1559P}
{Prentice}, S.~J., {Ashall}, C., {James}, P.~A., {et~al.} 2019, \mnras, 485,
  1559, \dodoi{10.1093/mnras/sty3399}

\bibitem[{{Renzo} {et~al.}(2019){Renzo}, {Zapartas}, {de Mink}, {G{\"o}tberg},
  {Justham}, {Farmer}, {Izzard}, {Toonen}, \& {Sana}}]{2019A&A...624A..66R}
{Renzo}, M., {Zapartas}, E., {de Mink}, S.~E., {et~al.} 2019, \aap, 624, A66,
  \dodoi{10.1051/0004-6361/201833297}

\bibitem[{{Ritchie} {et~al.}(2012){Ritchie}, {Stroud}, {Evans}, {Clark},
  {Hunter}, {Lennon}, {Langer}, \& {Smartt}}]{2012A&A...537A..29R}
{Ritchie}, B.~W., {Stroud}, V.~E., {Evans}, C.~J., {et~al.} 2012, \aap, 537,
  \dodoi{10.1051/0004-6361/201117716}

\bibitem[{{Romero-Ca{\~n}izales} {et~al.}(2014){Romero-Ca{\~n}izales},
  {Herrero-Illana}, {P{\'e}rez- Torres}, {Alberdi}, {Kankare}, {Bauer},
  {Ryder}, {Mattila}, {Conway}, {Beswick}, \& {Muxlow}}]{2014MNRAS.440.1067R}
{Romero-Ca{\~n}izales}, C., {Herrero-Illana}, R., {P{\'e}rez- Torres}, M.~A.,
  {et~al.} 2014, \mnras, 440, 1067, \dodoi{10.1093/mnras/stu430}

\bibitem[{{Roming} {et~al.}(2009){Roming}, {Pritchard}, {Brown}, {Holland},
  {Immler}, {Stockdale}, {Weiler}, {Panagia}, {Van Dyk}, {Hoversten}, {Milne},
  {Oates}, {Russell}, \& {Vandrevala}}]{2009ApJ...704L.118R}
{Roming}, P.~W.~A., {Pritchard}, T.~A., {Brown}, P.~J., {et~al.} 2009, \apj,
  704, L118, \dodoi{10.1088/0004-637X/704/2/L118}

\bibitem[{{Ryder} {et~al.}(2004){Ryder}, {Sadler}, {Subrahmanyan}, {Weiler},
  {Panagia}, \& {Stockdale}}]{2004MNRAS.349.1093R}
{Ryder}, S.~D., {Sadler}, E.~M., {Subrahmanyan}, R., {et~al.} 2004, \mnras,
  349, 1093, \dodoi{10.1111/j.1365-2966.2004.07589.x}

\bibitem[{{Ryder} {et~al.}(2018){Ryder}, {Van Dyk}, {Fox}, {Zapartas}, {de
  Mink}, {Smith}, {Brunsden}, {Azalee Bostroem}, {Filippenko}, {Shivvers}, \&
  {Zheng}}]{2018ApJ...856...83R}
{Ryder}, S.~D., {Van Dyk}, S.~D., {Fox}, O.~D., {et~al.} 2018, \apj, 856, 83,
  \dodoi{10.3847/1538-4357/aaaf1e}

\bibitem[{{Salpeter}(1955)}]{1955ApJ...121..161S}
{Salpeter}, E.~E. 1955, \apj, 121, 161, \dodoi{10.1086/145971}

\bibitem[{{Sana} {et~al.}(2012){Sana}, {de Mink}, {de Koter}, {Langer},
  {Evans}, {Gieles}, {Gosset}, {Izzard}, {Le Bouquin}, \&
  {Schneider}}]{2012Sci...337..444S}
{Sana}, H., {de Mink}, S.~E., {de Koter}, A., {et~al.} 2012, Science, 337, 444,
  \dodoi{10.1126/science.1223344}

\bibitem[{{Sapir} \& {Waxman}(2017)}]{2017ApJ...838..130S}
{Sapir}, N., \& {Waxman}, E. 2017, \apj, 838, 130,
  \dodoi{10.3847/1538-4357/aa64df}

\bibitem[{{Sato} {et~al.}(2020){Sato}, {Yoshida}, {Umeda}, {Nagataki}, {Ono},
  {Maeda}, {Hirai}, {Hughes}, {Williams}, \& {Maeda}}]{2020ApJ...893...49S}
{Sato}, T., {Yoshida}, T., {Umeda}, H., {et~al.} 2020, \apj, 893, 49,
  \dodoi{10.3847/1538-4357/ab822a}

\bibitem[{{Schure} {et~al.}(2008){Schure}, {Vink}, {Garc{\'\i}a-Segura}, \&
  {Achterberg}}]{2008ApJ...686..399S}
{Schure}, K.~M., {Vink}, J., {Garc{\'\i}a-Segura}, G., \& {Achterberg}, A.
  2008, \apj, 686, 399, \dodoi{10.1086/591432}

\bibitem[{{Smartt} {et~al.}(2009){Smartt}, {Eldridge}, {Crockett}, \&
  {Maund}}]{2009MNRAS.395.1409S}
{Smartt}, S.~J., {Eldridge}, J.~J., {Crockett}, R.~M., \& {Maund}, J.~R. 2009,
  \mnras, 395, 1409, \dodoi{10.1111/j.1365-2966.2009.14506.x}

\bibitem[{{Smith}(2014)}]{2014ARA&A..52..487S}
{Smith}, N. 2014, \araa, 52, 487, \dodoi{10.1146/annurev-astro-081913-040025}

\bibitem[{{Smith} {et~al.}(2011){Smith}, {Li}, {Filippenko}, \&
  {Chornock}}]{2011MNRAS.412.1522S}
{Smith}, N., {Li}, W., {Filippenko}, A.~V., \& {Chornock}, R. 2011, \mnras,
  412, 1522, \dodoi{10.1111/j.1365-2966.2011.17229.x}

\bibitem[{{Soderberg} {et~al.}(2006){Soderberg}, {Chevalier}, {Kulkarni}, \&
  {Frail}}]{2006ApJ...651.1005S}
{Soderberg}, A.~M., {Chevalier}, R.~A., {Kulkarni}, S.~R., \& {Frail}, D.~A.
  2006, \apj, 651, 1005, \dodoi{10.1086/507571}

\bibitem[{{Soker}(2017)}]{2017MNRAS.470L.102S}
{Soker}, N. 2017, \mnras, 470, L102, \dodoi{10.1093/mnrasl/slx089}

\bibitem[{{Sravan} {et~al.}(2019){Sravan}, {Marchant}, \&
  {Kalogera}}]{2019ApJ...885..130S}
{Sravan}, N., {Marchant}, P., \& {Kalogera}, V. 2019, \apj, 885, 130,
  \dodoi{10.3847/1538-4357/ab4ad7}

\bibitem[{{Strotjohann} {et~al.}(2015){Strotjohann}, {Ofek}, {Gal-Yam},
  {Sullivan}, {Kulkarni}, {Shaviv}, {Fremling}, {Kasliwal}, {Nugent}, {Cao},
  {Arcavi}, {Sollerman}, {Filippenko}, {Yaron}, {Laher}, \&
  {Surace}}]{2015ApJ...811..117S}
{Strotjohann}, N.~L., {Ofek}, E.~O., {Gal-Yam}, A., {et~al.} 2015, \apj, 811,
  117, \dodoi{10.1088/0004-637X/811/2/117}

\bibitem[{{Sundqvist} {et~al.}(2019){Sundqvist}, {Bj{\"o}rklund}, {Puls}, \&
  {Najarro}}]{2019A&A...632A.126S}
{Sundqvist}, J.~O., {Bj{\"o}rklund}, R., {Puls}, J., \& {Najarro}, F. 2019,
  \aap, 632, A126, \dodoi{10.1051/0004-6361/201936580}

\bibitem[{{Taddia} {et~al.}(2018){Taddia}, {Stritzinger}, {Bersten}, {Baron},
  {Burns}, {Contreras}, {Holmbo}, {Hsiao}, {Morrell}, {Phillips}, {Sollerman},
  \& {Suntzeff}}]{2018A&A...609A.136T}
{Taddia}, F., {Stritzinger}, M.~D., {Bersten}, M., {et~al.} 2018, \aap, 609,
  A136, \dodoi{10.1051/0004-6361/201730844}

\bibitem[{{Tartaglia} {et~al.}(2017){Tartaglia}, {Fraser}, {Sand}, {Valenti},
  {Smartt}, {McCully}, {Anderson}, {Arcavi}, {Elias-Rosa}, {Galbany},
  {Gal-Yam}, {Haislip}, {Hosseinzadeh}, {Howell}, {Inserra}, {Jha}, {Kankare},
  {Lundqvist}, {Maguire}, {Mattila}, {Reichart}, {Smith}, {Smith},
  {Stritzinger}, {Sullivan}, {Taddia}, \& {Tomasella}}]{2017ApJ...836L..12T}
{Tartaglia}, L., {Fraser}, M., {Sand}, D.~J., {et~al.} 2017, \apjl, 836, L12,
  \dodoi{10.3847/2041-8213/aa5c7f}

\bibitem[{{The LUVOIR Team}(2018)}]{2018arXiv180909668T}
{The LUVOIR Team}. 2018, arXiv e-prints, arXiv:1809.09668.
\newblock \doarXiv{1809.09668}

\bibitem[{{The LUVOIR Team}(2019)}]{2019arXiv191206219T}
---. 2019, arXiv e-prints, arXiv:1912.06219.
\newblock \doarXiv{1912.06219}

\bibitem[{{Tyson}(2002)}]{2002SPIE.4836...10T}
{Tyson}, J.~A. 2002, in \procspie, Vol. 4836, Survey and Other Telescope
  Technologies and Discoveries, ed. J.~A. {Tyson} \& S.~{Wolff}, 10--20,
  \dodoi{10.1117/12.456772}

\bibitem[{{Utrobin}(1994)}]{1994A&A...281L..89U}
{Utrobin}, V. 1994, \aap, 281, L89

\bibitem[{{Utrobin} \& {Chugai}(2009)}]{2009A&A...506..829U}
{Utrobin}, V.~P., \& {Chugai}, N.~N. 2009, \aap, 506, 829,
  \dodoi{10.1051/0004-6361/200912273}

\bibitem[{{Van Dyk} {et~al.}(2011){Van Dyk}, {Li}, {Cenko}, {Kasliwal},
  {Horesh}, {Ofek}, {Kraus}, {Silverman}, {Arcavi}, {Filippenko}, {Gal-Yam},
  {Quimby}, {Kulkarni}, {Yaron}, \& {Polishook}}]{2011ApJ...741L..28V}
{Van Dyk}, S.~D., {Li}, W., {Cenko}, S.~B., {et~al.} 2011, \apjl, 741, L28,
  \dodoi{10.1088/2041-8205/741/2/L28}

\bibitem[{{Van Dyk} {et~al.}(2014){Van Dyk}, {Zheng}, {Fox}, {Cenko}, {Clubb},
  {Filippenko}, {Foley}, {Miller}, {Smith}, {Kelly}, {Lee}, {Ben-Ami}, \&
  {Gal-Yam}}]{2014AJ....147...37V}
{Van Dyk}, S.~D., {Zheng}, W., {Fox}, O.~D., {et~al.} 2014, \aj, 147, 37,
  \dodoi{10.1088/0004-6256/147/2/37}

\bibitem[{{Van Dyk} {et~al.}(2018){Van Dyk}, {Zheng}, {Brink}, {Filippenko},
  {Milisavljevic}, {Andrews}, {Smith}, {Cignoni}, {Fox}, {Kelly}, {Adamo},
  {Yurus}, {Zhang}, \& {Kumar}}]{2018arXiv180301050V}
{Van Dyk}, S.~D., {Zheng}, W., {Brink}, T.~G., {et~al.} 2018, ArXiv e-prints.
\newblock \doarXiv{1803.01050}

\bibitem[{{van Loon}(2000)}]{2000A&A...354..125V}
{van Loon}, J.~T. 2000, \aap, 354, 125

\bibitem[{{van Loon} {et~al.}(2005){van Loon}, {Cioni}, {Zijlstra}, \&
  {Loup}}]{2005A&A...438..273V}
{van Loon}, J.~T., {Cioni}, M.-R.~L., {Zijlstra}, A.~A., \& {Loup}, C. 2005,
  \aap, 438, 273, \dodoi{10.1051/0004-6361:20042555}

\bibitem[{{Vink} {et~al.}(2001){Vink}, {de Koter}, \&
  {Lamers}}]{2001A&A...369..574V}
{Vink}, J.~S., {de Koter}, A., \& {Lamers}, H.~J.~G.~L.~M. 2001, \aap, 369,
  574, \dodoi{10.1051/0004-6361:20010127}

\bibitem[{{Weil} {et~al.}(2020){Weil}, {Fesen}, {Patnaude}, {Raymond},
  {Chevalier}, {Milisavljevic}, \& {Gerardy}}]{2020ApJ...891..116W}
{Weil}, K.~E., {Fesen}, R.~A., {Patnaude}, D.~J., {et~al.} 2020, \apj, 891,
  116, \dodoi{10.3847/1538-4357/ab76bf}

\bibitem[{{Wellons} {et~al.}(2012){Wellons}, {Soderberg}, \&
  {Chevalier}}]{2012ApJ...752...17W}
{Wellons}, S., {Soderberg}, A.~M., \& {Chevalier}, R.~A. 2012, \apj, 752, 17,
  \dodoi{10.1088/0004-637X/752/1/17}

\bibitem[{{Woosley} {et~al.}(1994){Woosley}, {Eastman}, {Weaver}, \&
  {Pinto}}]{1994ApJ...429..300W}
{Woosley}, S.~E., {Eastman}, R.~G., {Weaver}, T.~A., \& {Pinto}, P.~A. 1994,
  \apj, 429, 300, \dodoi{10.1086/174319}

\bibitem[{{Woosley} {et~al.}(1993){Woosley}, {Langer}, \&
  {Weaver}}]{1993ApJ...411..823W}
{Woosley}, S.~E., {Langer}, N., \& {Weaver}, T.~A. 1993, \apj, 411, 823,
  \dodoi{10.1086/172886}

\bibitem[{{Yoon} {et~al.}(2017){Yoon}, {Dessart}, \&
  {Clocchiatti}}]{2017ApJ...840...10Y}
{Yoon}, S.-C., {Dessart}, L., \& {Clocchiatti}, A. 2017, \apj, 840, 10,
  \dodoi{10.3847/1538-4357/aa6afe}

\bibitem[{{Yoon} {et~al.}(2006){Yoon}, {Langer}, \&
  {Norman}}]{2006A&A...460..199Y}
{Yoon}, S.-C., {Langer}, N., \& {Norman}, C. 2006, \aap, 460, 199,
  \dodoi{10.1051/0004-6361:20065912}

\bibitem[{{Yoon} {et~al.}(2010){Yoon}, {Woosley}, \&
  {Langer}}]{2010ApJ...725..940Y}
{Yoon}, S.-C., {Woosley}, S.~E., \& {Langer}, N. 2010, \apj, 725, 940,
  \dodoi{10.1088/0004-637X/725/1/940}

\bibitem[{{Young} {et~al.}(2006){Young}, {Fryer}, {Hungerford}, {Arnett},
  {Rockefeller}, {Timmes}, {Voit}, {Meakin}, \&
  {Eriksen}}]{2006ApJ...640..891Y}
{Young}, P.~A., {Fryer}, C.~L., {Hungerford}, A., {et~al.} 2006, \apj, 640,
  891, \dodoi{10.1086/500108}

\bibitem[{{Zhao} \& {Fuller}(2020)}]{2020MNRAS.tmp.1240Z}
{Zhao}, X., \& {Fuller}, J. 2020, \mnras, \dodoi{10.1093/mnras/staa1097}

\end{thebibliography}
\bibliographystyle{aasjournal}

\begin{longrotatetable}
\begin{deluxetable*}{lcccccccc}
\tablewidth{0pt}
\tabletypesize{\footnotesize}
\tablecaption{{{Properties of Type IIb SNe}} \label{t:SNIIbprops}}
\tablehead{
\colhead{} & \multicolumn{3}{c}{From Light Curves} & \multicolumn{4}{c}{From Pre-/Post-SN Photometry} & \colhead{From X-ray/Radio} 
\\
\cmidrule(lr){2-4} \cmidrule(lr){5-8} \cmidrule(lr){9-9} 
\colhead{SN} & \colhead{$M_{\rm env,1}/M_\odot$} & \colhead{$M_{\rm core,1}/M_\odot$} & \colhead{$R_{1}/R_\odot$} & \colhead{$\log_{10} (L_1/L_\odot)$} & \colhead{$\log_{10} (T_{\rm eff,1}/$K)} & \colhead{$\log_{10} (L_2/L_\odot)$} & \colhead{$\log_{10} (T_{\rm eff,2}/$K)} & \colhead{$\frac{\dot{M}_{\rm wind}/M_\sun{\rm yr}^{-1}}{v_{\rm wind}/{\rm km~s}^{-1}}$} 
}
\startdata
1993J & $0.15 - 0.4$ $^{1,2}$ & $2.8 - 6.0$ $^{1,3}$ & \nodata & $5.1 \pm 0.3$ $^{4*}$ & $3.63 \pm 0.05$ $^{4*}$ & $5.0 \pm 0.3$ $^{4*}$ & $4.3 \pm 0.1$ $^{4*}$ & $2-6\times10^{-6}$ $^5$ \\
2001gd & \nodata & \nodata & \nodata & \nodata & \nodata & \nodata & \nodata & $2-12\times10^{-6}$ $^6$ \\ 
2001ig & \nodata & \nodata & \nodata & \nodata & \nodata & $3.92\pm0.14$ $^7$ & $4.28-4.34$ $^7$ & $2.2\pm0.5\times10^{-6}$ $^{8}$ \\ 
2003bg & \nodata & \nodata & \nodata & \nodata & \nodata & \nodata & \nodata & $6.1-14 \times10^{-8}$ $^9$ \\ 
2008ax & \nodata & $\lesssim 5$ $^{10}$ & $30 - 50$ $^{10}$ & $4.42 - 5.3$ $^{10}$ & $3.88 - 4.3$ $^{10}$ & \nodata & \nodata & $9\pm3\times10^{-7}$ $^{11}$ \\ 
2009mg & \nodata & \nodata & \nodata & \nodata & \nodata & \nodata & \nodata & $< 1.5\times10^{-6}$ $^{12}$ \\ 
2010P & \nodata & \nodata & \nodata & \nodata & \nodata & \nodata & \nodata & $3-5.1\times10^{-6}$ $^{13}$ \\ 
2011dh & $\simeq0.1$ $^{14}$ & $3.06^{+0.68}_{-0.44}$ $^{15*}$ & $200 - 300$ $^{15}$ & $4.92 \pm 0.20$ $^{16}$ & $3.78 \pm 0.02$ $^{16}$ & \nodata & \nodata & $3\times10^{-8}$ $^{17}$ \\ 
2011ei & \nodata & \nodata & $\lesssim 1.5$ $^{18**}$ & \nodata & \nodata & \nodata & \nodata & $1.4\times10^{-8}$ $^{18}$ \\
2011fu & $\sim 0.3$ $^{19*}$ & \nodata & $\sim 450$ $^{19*}$ & \nodata & \nodata & \nodata & \nodata & \nodata \\
2011hs & $<0.5$  $^{20}$ & $3-4$ $^{20}$ & $500-600$ $^{20}$ & \nodata & \nodata & \nodata & \nodata & $2\pm0.6\times10^{-6}$ $^{20}$ \\ 
2012P & \nodata & $3.25^{+0.77}_{-0.56}$ $^{21}$ & \nodata & \nodata & \nodata & \nodata & \nodata & \nodata \\ 
2013df & $0.05-0.09$ $^{22**}$ & $2 - 3.6$ $^{22**}$ & $64-169$ $^{22**}$ & $4.94 \pm 0.06$ $^{23}$ & $3.63 \pm 0.01$ $^{23}$ & \nodata & \nodata & $10.5\pm3\times10^{-6}$ $^{24}$ \\ 
2015as & $\sim 0.1$ $^{25**}$ & \nodata & $\sim 7$ $^{25**}$ & \nodata & \nodata & \nodata & \nodata & \nodata \\
2016gkg & $0.01-0.4$ $^{26,27**}$ & \nodata & $40-340$ $^{26,27**}$ & $5.1^{+0.17}_{-0.19}$ $^{26*}$ & $3.86 \pm 0.05$ $^{26*}$ & \nodata & \nodata & \nodata \\ 
ZTF18aalrxas & $0.04-0.15$ $^{28**}$ & \nodata & $790-1050$ $^{28**}$ & \nodata & \nodata & \nodata & \nodata & \nodata 
\enddata
\tablecomments{\\ {\bf (1)} $M_{\rm env,1}$: Progenitor Hydrogen envelope mass; {\bf (2)} $M_{\rm core,1}$: Progenitor Helium core mass; {\bf (3)} $R_1$: Progenitor radius; {\bf (4)} $L_{1}$: Progenitor luminosity; {\bf (5)} $T_{\rm eff,1}$: Progenitor effective temperature; {\bf (6)} $L_{2}$: Companion luminosity; {\bf (7)} $T_{\rm eff,2}$: Companion effective temperature; {\bf (8)} $\dot{M}_{\rm wind}$: Progenitor mass loss rate; {\bf (9)} $v_{\rm wind}$: Progenitor wind velocity
\\ $^*$ Most recent value
\\ $^{**}$ Semi-analytic constraints}
\tablerefs{
$^1$ \cite{1994ApJ...429..300W}; $^2$ \cite{1996ApJ...456..811H}; $^3$ \cite{1993Natur.364..507N}; $^4$ \cite{2004Natur.427..129M};  $^5$ \cite{1996ApJ...461..993F};
$^6$ \cite{2005MNRAS.360.1055P};
$^7$ \cite{2018ApJ...856...83R}; $^8$ \cite{2004MNRAS.349.1093R}; 
$^9$ \cite{2006ApJ...651.1005S};
$^10$ \cite{2015ApJ...811..147F}; $^{11}$ \cite{2009ApJ...704L.118R}; 
$^{12}$ \cite{2012MNRAS.424.1297O};
$^{13}$ \cite{2014MNRAS.440.1067R};
$^{14}$ \cite{2012ApJ...757...31B}; $^{15}$ \cite{Ergon2015}; $^{16}$ \cite{2011ApJ...739L..37M}; $^{17}$ \cite{2012ApJ...750L..40K}; 
$^{18}$ \cite{2013ApJ...767...71M};
$^{19}$ \cite{2015MNRAS.454...95M};
$^{20}$ \cite{2014MNRAS.439.1807B};
$^{21}$ \cite{Fremling2016};
$^{22}$ \cite{2014MNRAS.445.1647M}; $^{23}$ \cite{2014AJ....147...37V}; $^{24}$ \cite{2016ApJ...818..111K};
$^{25}$ \cite{2018MNRAS.476.3611G};
$^{26}$ \cite{2018Natur.554..497B}; $^{27}$ \cite{2017ApJ...837L...2A};
$^{28}$ \cite{2019ApJ...878L...5F}
}
\end{deluxetable*}
\end{longrotatetable}

\end{document}